\documentclass[journal]{IEEEtran} 

\AtBeginDocument{%
  }

\usepackage{amsmath,amsfonts}
\usepackage{graphicx}
\usepackage{textcomp}
\usepackage{xcolor}

\usepackage{hyperref}
\usepackage{url}
\usepackage{tikz}
\usepackage{amsthm}
\usepackage{threeparttable}
\usepackage{mdframed}
\usepackage{calc}
\usepackage{algorithm, tabularx}
\usepackage[noend]{algpseudocode}
\usepackage{booktabs}
\usepackage{multirow}
\usepackage{xspace}
\usepackage{nicematrix}
\usepackage[tight,footnotesize]{subfigure}
\usepackage{pifont}
\usepackage{pdfpages}

\usepackage{enumitem}
\setlist[enumerate,1]{label=\arabic*.}

\usepackage{bibspacing}
\begin{document}
\sloppy

\input{commands}

\title{Secure and Efficient $L^p$-Norm Computation for Two-Party Learning Applications}

\author{Ali Arastehfard, Weiran Liu, Joshua Lee, Bingyu Liu, Xuegang Ban,~\IEEEmembership{Member,~IEEE}, and Yuan Hong,~\IEEEmembership{Senior Member,~IEEE}
	\thanks{Ali Arastehfard and Yuan Hong are with the School of Computing, University of Connecticut, USA (Email: ali.arastehfard@uconn.edu, yuan.hong@uconn.edu). Weiran Liu is with the Alibaba Group (Email: weiran.lwr@alibaba-inc.com). Joshua Lee is with the University of California Santa Barbara, USA (Email: jlee246@ucsb.edu). Bingyu Liu is with the School of Computing \& Data Science, Wentworth Institute of Technology, USA (Email: liub@wit.edu). Xuegang Ban is with the University of Washington, USA (Email: banx@uw.edu).}}

\markboth{}{}

\maketitle

\begin{abstract}
	Secure norm computation is becoming increasingly important in many real-world learning applications. However, existing cryptographic systems often lack a general framework for securely computing the $L^p$-norm over private inputs held by different parties. These systems often treat secure norm computation as a black-box process, neglecting to design tailored cryptographic protocols that optimize performance. Moreover, they predominantly focus on the $L^2$-norm, paying little attention to other popular $L^p$-norms, such as $L^1$ and $L^\infty$, which are commonly used in practice, such as machine learning tasks and location-based services.

	To our best knowledge, we propose the first comprehensive framework for secure two-party $L^p$-norm computations ($L^1$, $L^2$, and $L^\infty$), denoted as \mbox{Crypto-$L^p$}, designed to be versatile across various applications. We have designed, implemented, and thoroughly evaluated our framework across a wide range of benchmarking applications, state-of-the-art (SOTA) cryptographic protocols, and real-world datasets to validate its effectiveness and practical applicability. In summary, \mbox{Crypto-$L^p$} outperforms prior works on secure $L^p$-norm computation, achieving $82\times$, $271\times$, and $42\times$ improvements in runtime while reducing communication overhead by $36\times$, $4\times$, and $21\times$ for $p=1$, $2$, and $\infty$, respectively. Furthermore, we take the first step in adapting our Crypto-$L^p$ framework for secure machine learning inference, reducing communication costs by $3\times$ compared to SOTA systems while maintaining comparable runtime and accuracy. Code for \sys is available at \url{https://github.com/datasec-lab/cryptoLp}.
\end{abstract}
\section{Introduction} \label{sec:intro}
Since the foundational theoretical feasibility studies \cite{goldreich_how_1987, yao_protocols_1982}, secure multiparty computation (MPC) has undergone significant advancements, resulting in more efficient methods for multiple parties to collaboratively compute a function while preserving the privacy of their individual inputs. Secure two-party computation (2PC) is a major subfield within MPC, enabling two parties, $P_0$ and $P_1$, to jointly compute a function $f$ using their private inputs $x_0$ and $x_1$, respectively. This process ensures that no additional information is disclosed beyond the output of $f(x_0, x_1)$.

General solutions for secure 2PC (and also MPC) can be used to compute any discrete function that can be represented as a fixed-size circuit. However, for specific functionalities, there may be custom solutions that are much more efficient than the best generic ones. Private set intersection (PSI) is such a famous example \cite{DBLP:conf/sigecom/HubermanFH99,DBLP:conf/ndss/HuangEK12,DBLP:conf/ccs/KolesnikovKRT16}. The research community has shifted its focus to finding efficient methods for evaluating other specific functionalities. For example, Rathee et al.~\cite{rathee_sirnn_2021} focus on custom solutions for secure recurrent neural network (RNN) inferences by optimizing involved secure mathematical functionalities such as \emph{Sigmoid}, \emph{Tanh}, and \emph{Square Root}. Some of these functionalities are not only used in RNN, but are also widely used for other applications. Specifically, \emph{Square Root} is the major building block for computing the $L^2$-norm. The general $L^p$-norms ($L^1$, $L^2$ and $L^\infty$) are extensively used as distance metrics in a wide variety of applications, including machine learning (ML) \cite{mohassel_practical_2019}, biometric matching \cite{demmler_aby_2015, bringer_gshade_2014}, and location-based services (LBS) \cite{han_location_2020}. They frequently require performing computations across multiple parties with private inputs, and it is desirable to efficiently address the privacy concerns for the $L^p$-norm computations ubiquituously executed in such applications.


Despite the prevalent use of secure norm computations in several applications (including secure ML tasks \cite{mohassel_practical_2019,jaschke_unsupervised_2019}), there remains a consistent lack of a comprehensive framework for the secure computation of popular $L^p$-norms that could be broadly applied to various contexts. Numerous works in the field specialize in application-oriented research, where instances of constructions for secure norm computations can be observed \cite{han_location_2020,mohassel_practical_2019,demmler_aby_2015,bringer_gshade_2014,jaschke_unsupervised_2019}. However, a common limitation among these studies is that they often do not include the secure protocols used for their norm computations, treating them instead as a black box, or tailoring their constructions so specifically to a particular application that adapting them to other contexts becomes exceedingly challenging. For example, Demmler et al. (ABY)~\cite{demmler_aby_2015} and Bringer et al.~\cite{bringer_gshade_2014} utilize secure Euclidean distance ($L^2$-norm) as a distance metric in private biometric identification applications; Mohassel et al.~\cite{mohassel_practical_2019} and J{\"a}schke et al.~\cite{jaschke_unsupervised_2019} employ secure $L^2$-norm and $L^1$-norm for privacy-preserving clustering, respectively. While Han et al.~\cite{han_location_2020} demonstrates the use of all three norms, their construction is highly specific to the spatial crowdsourcing application and relies on the existence of a trusted third party, which limits their adaptability to other applications that require standard secure 2PC solutions.

Furthermore, over the past decade, extensive research has been conducted on the secure computation of ML tasks, especially secure deep neural network (DNN) inferences \cite{rathee_sirnn_2021,rathee_cryptflow2_2020,demmler_aby_2015,huang_cheetah_2022,juvekar_gazelle_2018,lehmkuhl_muse_2021,XieLH21,mishra_delphi_2020,mohassel_aby3_2018,mohassel_secureml_2017,rathee_secfloat_2022}. These advancements have led to significant optimizations in general-purpose secure primitives, such as the Millionaires' Protocol \cite{rathee_cryptflow2_2020}, Oblivious Transfer (OT)-based multiplication \cite{rathee_sirnn_2021}, and the rotation-free homomorphic scheme \cite{huang_cheetah_2022}. However, these works have generally tailored their protocols specifically for secure inference tasks, posing significant challenges in adapting them for other purposes like secure norm computation. In addition, although these inference libraries share similar building blocks, using them for tasks that involve secure norm computations (e.g., clustering, LBS) can be very frustrating due to the lack of an efficient cryptographic $L^p$-norm framework.

\begin{figure}[!h]
	\centering
	\includegraphics[width=1\columnwidth]{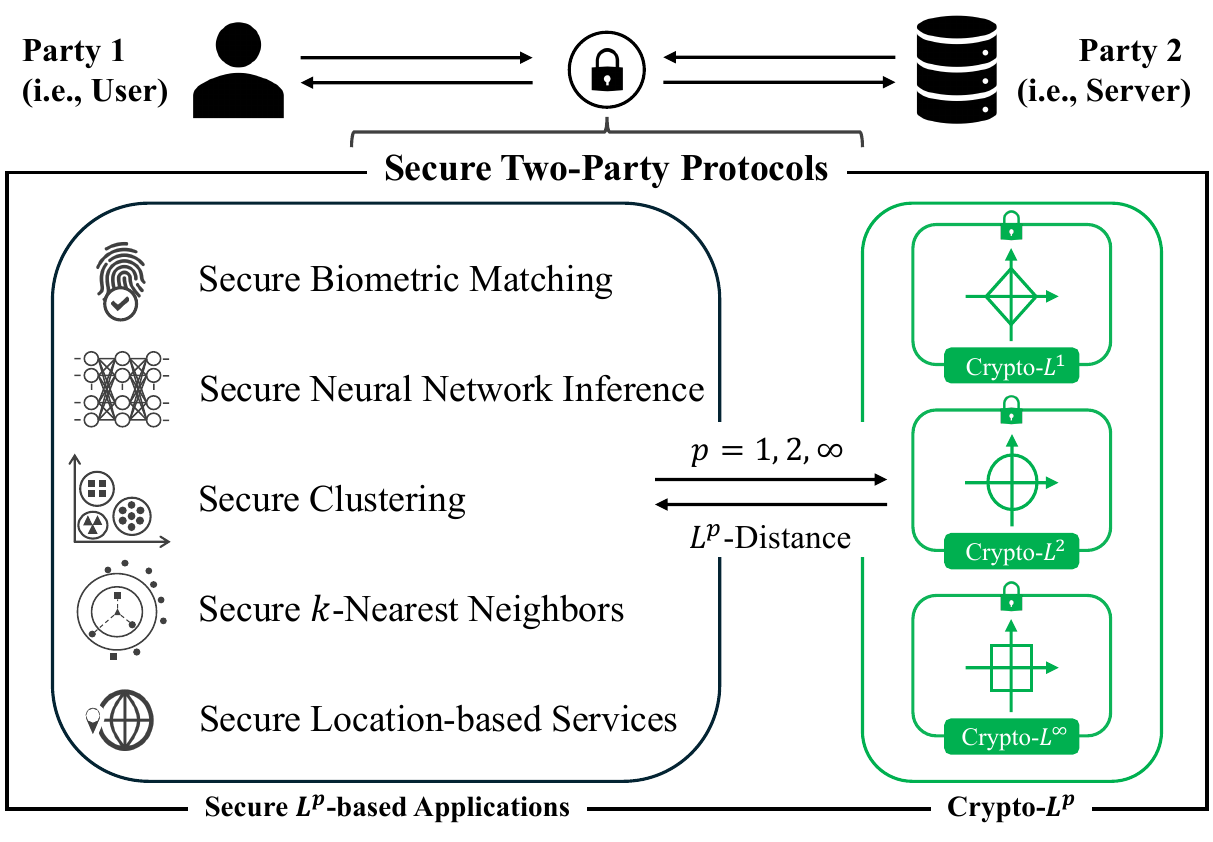}
	\caption{Representative applications of \sys. The listed applications are not exhaustive and merely serve as popular representatives for $L^p$-based computations.}
	\label{fig:framework_overview}
\end{figure}

To address these critical deficiencies, we propose, to the best of our knowledge, the first comprehensive framework for generalized secure norm computations, called \sys. Designed to enable the private computation of distance metrics across various applications, such as secure clustering, biometric matching, and LBS, the framework optimizes new cryptographic tools for $L^p$-norms in the secure 2PC setting. Figure \ref{fig:framework_overview} provides a high-level overview of the \sys framework and its integration with some representative secure $L^p$-based applications.

\subsection{Representative Applications for \sys{}} \label{subsec:applications}
Specifically, we illustrate some representative applications where \sys serves as a fundamental building block.

\vspace{0.05in}

\noindent\textbf{Secure 2PC Inference.}
We made a key observation when integrating \sys{} into DNN inference (Appendix~\ref{appendix:cnn}): replacing the convolution operation in CNNs with the $L^1$-norm shows strong potential in the secure 2PC setting. This substitution can significantly reduce the cost of secure two-party inference compared to SOTA frameworks~\cite{rathee_sirnn_2021,rathee_cryptflow2_2020,demmler_aby_2015,huang_cheetah_2022,juvekar_gazelle_2018,lehmkuhl_muse_2021,mishra_delphi_2020,mohassel_aby3_2018,mohassel_secureml_2017,rathee_secfloat_2022}. Prior works~\cite{addernet,xu_kernel_2020,you_shiftaddnet_2020} in plaintext settings support this substitution, showing that the $L^1$-norm offers similar accuracy to convolution while reducing multiplications—beneficial for energy-constrained devices~\cite{addernet}.

Motivated by this, we conducted extensive experiments to assess its viability in 2PC. Our results show that the $L^1$-norm (Adder operation~\cite{addernet}) is an effective drop-in replacement for convolution, particularly in high-dimensional settings. Notably, it reduces communication by up to $3\times$ while maintaining similar runtime and accuracy (see Section~\ref{eval:addernet}).

Given this novel application of \sys{} to secure ML inference\footnote{While secure norm computations have been studied in contexts such as clustering and biometric matching, \sys{} offers more efficient constructions. Its adaptation to secure ML inference, however, is novel.}, Section~\ref{sec:secure_addernet} details our secure $L^p$-norm protocols—specifically the $L^1$-norm—for ML inference tasks in the 2PC setting.

\vspace{0.05in}

\noindent\textbf{Secure Clustering.} Clustering is an unsupervised learning task that groups data points such that points in the same cluster are more similar to each other than to those in other clusters. In the secure 2PC setting, two parties jointly execute the clustering algorithm on horizontally or vertically partitioned data~\cite{mohassel_practical_2019}, without revealing anything beyond the final centroids. Popular clustering algorithms like $K$-means minimize a cost function based on distances between points and their cluster centroids. Mohassel et al.~\cite{mohassel_practical_2019} proposed a secure clustering protocol that protects intermediate centroid values and supports Manhattan ($L^1$), Euclidean ($L^2$), and Chebyshev ($L^\infty$) distances. In Section~\ref{sec:evals}, we evaluate and compare our more efficient secure norm protocols, which serve as distance metrics for secure clustering.

\vspace{0.05in}

\noindent\textbf{Secure Biometric Matching.} In secure 2PC biometric matching applications, one party wants to determine whether its biometric sample (e.g., fingerprint, face) matches any biometric samples stored in a database held by another party. A fundamental building block of these protocols is the computation of the squared Euclidean distance between the query and all biometric samples in the database, followed by identifying the minimum among these distances \cite{demmler_aby_2015}. To achieve this securely, one party holds their $n$-dimensional biometric vector $x$, and the other party holds a database $D$ consisting of $m$ other $n$-dimensional biometric vectors.
They aim to securely compute $\underset{i}{\min}(\sum\limits_{j=1}^{n} (D_{i, j} - x_j)^2)$ for $i \in \{1,\dots, m\}$ without disclosing their data. Later, we will construct functionalities to securely compute the Euclidean distance and the minimum\footnote{We develop secure protocols for the maximum functionality for our $L^{\infty}$. The minimum functionality can be implemented similarly.} of vectors that facilitate this computation securely.

\vspace{0.05in}

\noindent\textbf{Secure Location-based Services.} Location-based services (LBS) are applications that utilize user location data to provide tailored services. For instance, platforms like Yelp use location information to identify the nearest restaurants to a user's current location. These proximity-oriented applications inherently rely on distance metrics to determine the nearest routes or entities to a user. Manhattan, Euclidean, and Chebyshev distances are popular metrics, each suited to different scenarios \cite{han_location_2020}. In a secure two-party setting, we assume that the user's location remains confidential and that the server does not disclose any data from its dataset, except for the search results (e.g., $k$ nearest restaurants, assigned drivers). Accordingly, we evaluate and compare our efficient secure norm protocols with related works in the context of secure LBS in Section~\ref{sec:evals}.

\subsection{Our Contributions}

Our main contributions are summarized as follows:

\begin{itemize}[leftmargin=*]
	\item \textbf{A Unified Framework for Cryptographic Norms:} We present \sys{}, the first cryptographic norm framework for secure and efficient $L^p$-norm computations across a wide range of learning applications in the 2PC setting. \sys{} serves as a versatile toolkit for privacy-preserving data analysis.

	      \vspace{0.05in}

	\item \textbf{Efficient and Modular Cryptographic Building Blocks:} \sys{} introduces optimized primitives tailored for secure $L^p$-norms. For example, our custom absolute value multiplexer reduces both communication and computation by $2\times$ compared to naively composed alternatives.

	      \vspace{0.05in}

	\item \textbf{First $L^1$-Norm-Based 2PC Neural Network:} We observe that replacing convolutions in CNNs with $L^1$-norms in a secure 2PC setting can reduce communication by up to $3\times$, with minimal impact on runtime and accuracy. To our knowledge, \sys{} is the first to explore this substitution in 2PC and evaluate its trade-offs, opening new directions for PPML.

	      \vspace{0.05in}

	\item \textbf{Extensive Evaluation and Superior Performance:} We validate \sys{} through comprehensive experiments on both core building blocks and downstream tasks. Compared to state-of-the-art, \sys{} improves runtime by up to $82\times$, $271\times$, and $42\times$, and reduces communication by $36\times$, $4\times$, and $21\times$ for $p=1$, $2$, and $\infty$, respectively.
\end{itemize}

\section{Preliminaries}

\subsection{Notations}
We first establish the notations that will be used throughout the paper. Lowercase letters, such as $x$, denote $n$-dimensional vectors of $\ell$-bit integers. To represent $\ell$-bit integers, we set $n=1$ and treat them as one-dimensional vectors. For instance, $x \in \mathbb{Z}_L$ indicates that $x$ is an $\ell$-bit integer, while $x \in \mathbb{Z}_L^n$ indicates that $x$ is an $n$-dimensional vector consisting of $\ell$-bit integers, where $L$ represents $2^\ell$ (see Section \ref{subsec:primitives}). Uppercase letters, such as $X$, denote matrices and tensors. The  computational security parameter $\lambda$ is set to be $128$. Additionally, $1\{P\}$ denotes that for a condition $P$, if $P$ is true, it returns 1; otherwise, it returns 0. We write $\overline{b}$ for the binary complement of $b$, i.e., $1 - b$.

\subsection{Threat Model and Security} \label{subsec:threat_model}
We adopt the semi-honest (passive) adversary model, where we assume a computationally bounded adversary $\mathcal{A}$ who attempts to learn additional information from the messages observed during protocol execution, yet is not permitted to deviate from the protocol. As in prior works \cite{rathee_cryptflow2_2020, huang_cheetah_2022, rathee_sirnn_2021, rathee_secfloat_2022}, we present our protocols using the hybrid model. This model simulates real interactions by substituting the execution of sub-protocols with calls to their respective trusted functionalities $\mathcal{F}$. For example, assume that two parties, $P_0$ and $P_1$, wish to jointly compute a functionality $f$ while securing their inputs. In this simulation, $P_0$ and $P_1$ provide their inputs to the trusted functionality $\mathcal{F}$, which computes $f$ and returns the result back to the parties. In such a scenario, we say that any protocol utilizing $\mathcal{F}$ operates within the $\mathcal{F}$-hybrid model.

\subsection{Cryptographic Primitives} \label{subsec:primitives}

Before introducing our protocol design, we first review some necessary cryptographic primitives.

\vspace{0.05in}

\noindent\textbf{Secret Sharing}. 
We employ 2-out-of-2 secret sharing schemes~\cite{shamir_how_1979, blakley_safeguarding_1979, demmler_aby_2015} between two parties, $P_0$ and $P_1$, over the ring $\mathbb{Z}_{L}$ for $\ell$-bit arithmetic messages and $\mathbb{Z}_2$ for Boolean messages. An arithmetic value $x \in \mathbb{Z}_{L}$ is shared as $\ashare{x}{L}_b$ for $b \in \{0,1\}$ such that $x = (\ashare{x}{L}_0 + \ashare{x}{L}_1) \bmod 2^{\ell}$. Similarly, a Boolean bit $s \in \mathbb{Z}_2$ is shared as $\bshare{s}_b$. These shares reveal no information about the secret, as each share is uniformly random and independent of the underlying value.

\vspace{0.05in}

\noindent\textbf{Oblivious Transfer}.
We adopt oblivious transfer (OT) as a fundamental primitive in our work. Specifically, we utilize $\binom{2}{1}\text{-OT}_{\ell}$, where the sender inputs two $\ell$-bit strings $(s_0,s_1)$ and the receiver inputs a selection bit $b \in \{0,1\}$, thereby obliviously learning $s_b$. By the end of the protocol, the receiver gains no information about $s_{\bar{b}}$ and the sender learns no information about $b$. This extends to $m \times \text{OT}_{\ell}$ over $m$ input pairs $(s^i_0, s^i_1)$.

On its own, 2-round OT has been shown to be equivalent to 2-round key exchange \cite{gertner_relationship_2000}, which requires costly public-key cryptography. However, OT extensions \iknp{}---often referred to as IKNP-style OTs---enable the implementation of $m \times \text{OT}_{\ell}$ with only a small number of actual OTs, known as base OTs. These base OTs are then extended through inexpensive symmetric cryptographic operations and a constant number of rounds.

There are different variations of OT that can help further improve its efficiency. In Random OT (ROT) \cite{asharov_more_2013}, the problem of $\binom{2}{1}\text{-OT}_{\ell}$ is reduced to $\binom{2}{1}\text{-ROT}_{\ell}$ through a preprocessing step that invokes the OT functionality on two random strings. Formally, the receiver inputs a choice bit $b$, while the sender has no inputs. The ROT functionality provides the sender with two random $\ell$-bit strings $(s_0,s_1)$, and the receiver learns $s_b$. This idea of having preprocessed OTs was first introduced by Beaver \cite{beaver_precomputing_1995}. Correlated OT (COT) \cite{asharov_more_2013} builds upon ROT by modifying the preprocessing phase. Unlike in ROT, the sender in $\binom{2}{1}\text{-COT}_{\ell}$ inputs a correlation function $f_{\Delta}(\cdot)$, and obtains a random string $s_0$ and a correlated string $s_1 = f_{\Delta}(s_0)$. The receiver, after inputting their choice bit $b$, receives $s_b$, which is derived based on the predefined correlation.

\vspace{0.05in}

\noindent\textbf{Multiplication via OT}. 
Multiplication is generally performed using pre-computed arithmetic multiplication triples~\cite{feigenbaum_efficient_1992}. To multiply two shared values $\ashare{x}{L}$ and $\ashare{y}{L}$, prior works~\cite{demmler_aby_2015, mohassel_secureml_2017} precompute triples $(\ashare{a}{L}, \ashare{b}{L}, \ashare{c}{L})$ such that $c = a \cdot b \mod 2^{\ell}$. Each party $P_i$ computes $\ashare{e}{L}_i = \ashare{x}{L}_i - \ashare{a}{L}_i$ and $\ashare{f}{L}_i = \ashare{y}{L}_i - \ashare{b}{L}_i$, and sets $\ashare{z}{L}_i = -i \cdot e \cdot f + f \cdot \ashare{x}{L}_i + e \cdot \ashare{y}{L}_i + \ashare{c}{L}_i$.

One can generate the multiplication triples using OT \cite{demmler_aby_2015, goos_two_1999}. To generate an arithmetic multiplication triple $\ashare{c}{L} = \ashare{a}{L} \cdot \ashare{b}{L}$, consider \vspace{1pt} that $\ashare{a}{L} \cdot \ashare{b}{L} = (\ashare{a}{L}_0 + \ashare{a}{L}_1) \cdot (\ashare{b}{L}_0 + \ashare{b}{L}_1) = \ashare{a}{L}_0 \cdot \ashare{b}{L}_0 + \ashare{a}{L}_0 \cdot \ashare{b}{L}_1 \vspace{1pt} + \ashare{a}{L}_1 \cdot \ashare{b}{L}_0 + \ashare{a}{L}_1 \cdot \ashare{b}{L}_1$. While parties $P_0$ and $P_1$ can compute $\ashare{a}{L}_0 \cdot \ashare{b}{L}_0$ and $\ashare{a}{L}_1 \cdot \ashare{b}{L}_1$ locally, the non-local products $\ashare{a}{L}_0 \cdot \ashare{b}{L}_1$ and $\ashare{a}{L}_1 \cdot \ashare{b}{L}_0$ require the use of OT \cite{demmler_aby_2015, goos_two_1999}. In ABY \cite{demmler_aby_2015}, parties engage in ${\ell} \times \binom{2}{1}\text{-COT}_{\ell}$ to compute $\ashare{u}{L} = \ashare{\ashare{a}{L}_0 \cdot \ashare{b}{L}_1}{L}$, where in the $j$-th $\binom{2}{1}\text{-COT}_{\ell}$, $P_0$ is the sender with the correlation $\rho(s_{j}) = (\ashare{a}{L}_0 \cdot 2^j - s_j) \bmod 2^{\ell}$ and $P_1$ is the receiver with input \vspace{1pt} $\ashare{b}{L}_1[j]$ (the $j$-th bit of $\ashare{b}{L}_1$). In the $j$-th iteration, $P_1$ learns $t_{i,\ashare{b}{L}_1[j]} = (\ashare{b}{L}_1[j] \cdot \ashare{a}{L}_0 \cdot 2^j - s_j) \bmod 2^{\ell}$. Subsequently, $P_0$ sets $\ashare{u}{L}_0 = (\sum_{j=1}^{\ell} s_j) \bmod 2^{\ell}$, and $P_1$ sets $\ashare{u}{L}_1 = (\sum_{j=1}^{\ell} t_{i,\ashare{b}{L}_1[j]}) \bmod 2^{\ell}$ \vspace{1pt}. Similarly, parties compute $\ashare{v}{L} = \ashare{\ashare{a}{L}_1 \cdot \ashare{b}{L}_0}{L} \vspace{1pt}$, and at the end of the protocol, $P_i$ sets $\ashare{c}{L}_i = \ashare{a}{L}_i \cdot \ashare{b}{L}_i + \ashare{u}{L}_i + \ashare{v}{L}_i$.

We can further optimize the multiplication \vspace{1pt} protocol by substituting $\binom{2}{1}\text{-COT}_{\ell}$ with the $\binom{N}{1}\text{-COT}_{\ell}$ scheme of \cite{dessouky_pushing_2018,mohassel_practical_2019}. At a high level, in this variation, during the precomputation of multiplication triples, $P_1$ employs an $N$-base representation instead of the binary representation of their input $b$, rewriting \vspace{1pt} $b = \sum_{j=1}^{\ell / \log(N)} N^j b_j$. Subsequently, $P_0$ and $P_1$ invoke $\lceil {\ell} / {\log(N)} \rceil$ instances of $\binom{N}{1}\text{-COT}_{\ell}$ \vspace{1pt} to obtain arithmetic shares of each $N^j b_j a$, where $P_0$, $P_1$ holds $a$ and $b$, respectively.

\vspace{0.05in}

\noindent\textbf{Millionaires' Problem} ($\mathcal{F}_{\text{Mill}}^{\ell-1}$). In Yao's Millionaires' problem, two parties $P_0$ and $P_1$ who hold $\ell$-bit integers $x$ and $y$, respectively, wish to learn Boolean shares of $1\{x < y\}$. In this paper, our building blocks are based on the Millionaires' protocol of Rathee et al.~\cite{rathee_cryptflow2_2020}.

The idea in ~\cite{rathee_cryptflow2_2020} is primarily based on the observation that the comparison between any $\ell$-bit integers $x$ and $y$ can be broken down into two separate comparisons and one equality test. Specifically, given that $x = x_0x_1$ and $y = y_0y_1$, the following equation holds:
\begin{equation}\label{eq:cmp}
	1\{ x < y \} = 1\{ x_1 < y_1 \} \oplus (1\{ x_1 = y_1 \} \wedge 1\{ x_0 < y_0 \})
\end{equation}
Per the above observation, the protocol in \cite{rathee_cryptflow2_2020} is as follows:
\begin{enumerate}[leftmargin=*]
	\item Each party $P_b$ parses its own $\ell$-bit input into smaller chunks of $m$-bit integers. Let $M = 2^m$, and $x_j$, $y_j$ denote the $j$-th block in integers \vspace{1pt} $x$ and $y$, respectively. The two parties invoke two instances of $\binom{M}{1}\text{-OT}_{1}$ to learn Boolean shares of $\text{lt}_j = 1\{x_j < y_j\}$ and $\text{eq}_j = 1\{x_j = y_j\}$. \label{step:parse}
	\item Let $\func{\text{AND}}{}$ denote the functionality that accepts two Boolean shares and returns Boolean shares of their $\text{AND}$. The two parties use Eq.~\ref{eq:cmp} to construct a perfect binary tree. At the leaf level, initial Boolean shares of inequalities ($\text{lt}_j$) and equalities ($\text{eq}_j$) for each chunk are combined using $\func{\text{AND}}{}$ to construct the shares of inequalities and equalities at higher levels. This process is repeated progressively up the tree, and the value of the inequality at the root provides the final output. \label{step:combine}
\end{enumerate}

As mentioned in \cite{rathee_cryptflow2_2020}, the idea of having a perfect binary tree requires that $m$ divides $\ell$ and that $\ell/m$ is a power of $2$. In cases where $\ell/m$ is not a power of $2$, Rathee et al.~\cite{rathee_cryptflow2_2020} addresses this by constructing the largest possible perfect binary trees and connecting their roots using Eq.~\ref{eq:cmp}. For ease of exposition, let us assume that $\ell/m$ is a power of $2$. Under this assumption, it is straightforward to see that the number of $\text{AND}$ operations invoked in the protocol equals the number of nodes in the tree (excluding the root), which is $2 \times (\ell/m - 1)$. This can further be optimized by eliminating the redundant $\text{AND}$ operations in the rightmost branch of each tree level (stored from MSB to LSB). With this optimization, the number of $\text{AND}$ operations can be reduced to a minimum of $2 \times (\ell/m - 1)$ minus the depth of the tree (excluding the root level), which is $2 \times (\ell/m - 1) - \log(\ell/m)$.


\section{Building Blocks} \label{sec:core}

\begin{figure*}[htbp]
	\centering
	\includegraphics[width=\textwidth]{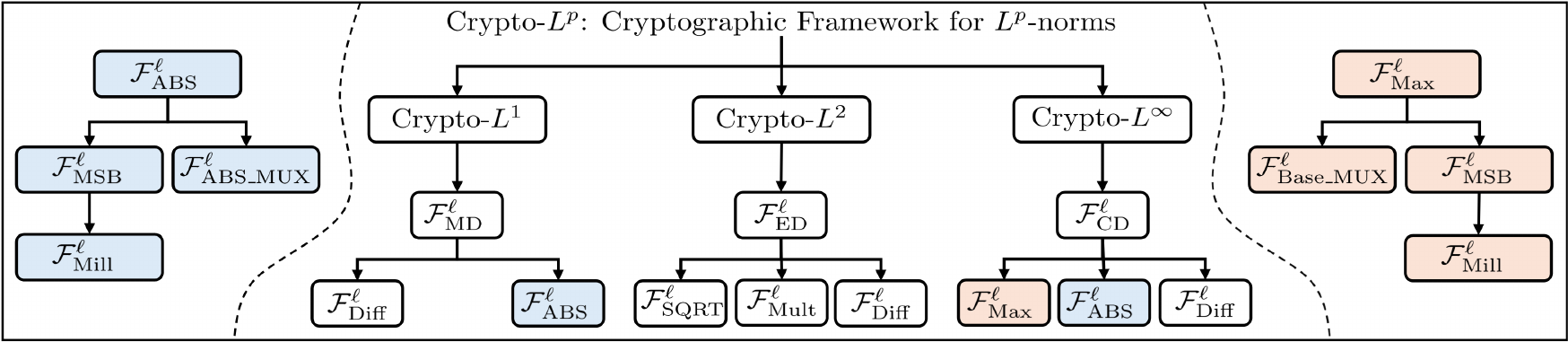}
	\caption{Overview of the building blocks in \sys ($p=1, 2$ and $\infty$).}
	\label{fig:building_blocks}
\end{figure*}

The core of \sys involves essential functionalities for secure norm computations (see Figure~\ref{fig:building_blocks}).
In this section, we review and optimize these core cryptographic building blocks.
The communication complexities for each of these building blocks are summarized in Table~\ref{tab:algs-summary}.

\subsection{Secure Absolute Value Computation $|x|$}
\label{sec:secure_absolute}

In this section, we present our protocol designed to compute the absolute value of an $\ell$-bit integer $x$, denoted $|x|$. The operation of computing the absolute value fundamentally involves checking the sign of $x$ and, if negative, negating it to obtain a positive value.

We design $\mathcal{F}_{\text{ABS}}^\ell$ (Alg.~\ref{alg:abs}) to realize this functionality. Specifically, $\mathcal{F}_{\text{ABS}}^\ell$ begins by invoking $\mathcal{F}_{\text{MSB}}^\ell$, which distributes shares of $x$'s most significant bit (MSB) to the parties $P_0$ and $P_1$, represented as $\bshare{\mathrm{msb}}_0$ and $\bshare{\mathrm{msb}}_1$, respectively. These shares then serve as selector bits in the subsequent ABS Multiplexer functionality, $\mathcal{F}_{\text{ABS\_MUX}}^\ell$, allowing the parties to obliviously select either $x$ or $-x$ based on the value of $\mathrm{msb}$. Upon completion of $\mathcal{F}_{\text{ABS}}^\ell$, each party will hold shares of $|x|$, denoted as $\ashare{\mathrm{abs}}{L}_0$ and $\ashare{\mathrm{abs}}{L}_1$. Next, we focus on designing $\mathcal{F}_{\text{ABS\_MUX}}^\ell$ and $\mathcal{F}_{\text{MSB}}^\ell$ that are invoked by $\mathcal{F}_{\text{ABS}}^\ell$.

\begin{algorithm}
	\small
	\caption{Absolute Value, $\mathcal{F}^{\ell}_{\text{ABS}}$}\label{alg:abs}
	\begin{algorithmic}[1]
		\Input{For $b \in \{0,1\}$, $P_b$ holds $\ashare{x}{L}_b$.}
		\Output{For $b \in \{0, 1\}$, $P_b$ learns $\ashare{\mathrm{abs}}{L}_b$ s.t. $\mathrm{abs} = |x|$.}
		\State \multiline{%
		$P_0$ and $P_1$ invoke an instance of $\mathcal{F}^{\ell}_{\text{MSB}}$. For $b \in \{0,1\}$, $P_b$ inputs $\ashare{x}{L}_b$ and learns $\bshare{\mathrm{msb}}_b$.}
		\State \multiline{%
		$P_0$ and $P_1$ invoke an instance of $\mathcal{F}^{\ell}_{\text{ABS\_MUX}}$. For $b \in \{0,1\}$, $P_b$ inputs ($\bshare{\mathrm{msb}}_b$, $\ashare{x}{L}_b$) and learns $\ashare{\mathrm{abs}}{L}_b$.}
		\State For $b \in \{0,1\}$, $P_b$ outputs $\ashare{\mathrm{abs}}{L}_b$.
	\end{algorithmic}
\end{algorithm}

\vspace{0.05in}

\noindent\textbf{Multiplexer for Absolute Value}. \label{sec:multiplexer_abs}
What we require in $\mathcal{F}_{\text{ABS\_MUX}}^{\ell}$ is to select between $x$ and $-x$, where $x$ is an $\ell$-bit integer. To realize this, we use a multiplexer protocol similar to \cite{rathee_cryptflow2_2020}, denoted by the base multiplexer $\mathcal{F}_{\text{Base\_MUX}}^{\ell}$ (Alg.~\ref{alg:base_mux} Appendix~\ref{appendix:base_mux}), which selects between $x$ and $0$ (i.e., enables or disables the value). A naive realization of $\mathcal{F}_{\text{ABS\_MUX}}^{\ell}$ is to invoke the base multiplexer twice with inputs $(s, (x, 0))$ and $(\bar{s}, (-x, 0))$, and then add the two outputs. If $s=0$, the first call outputs $x$ and the second outputs $0$, so $x+0=x$; if $s=1$, they output $0$ and $-x$, so $0+(-x)=-x$. This realizes $\mathcal{F}_{\text{ABS\_MUX}}^{\ell}$.

We observe that these two base multiplexers can be combined using the relation $\mathcal{F}_{\text{ABS\_MUX}}^{\ell}(x, s) = x - 2 \cdot \mathcal{F}_{\text{Base\_MUX}}^{\ell}(x, s)$. This reduces computation and communication cost by half, as it halves the number of COTs. Details are provided in Algorithm~\ref{alg:abs_mux}, and Appendix~\ref{sec:correct_abs_mux} proves correctness.

\begin{algorithm}
	\small
	\caption{ABS Multiplexer, $\mathcal{F}^{\ell}_{\text{ABS\_MUX}}$} \label{alg:abs_mux}
	\begin{algorithmic}[1]
		\Input{For $b \in \{0,1\}$, $P_b$ holds $\ashare{x}{L}_b$ and $\bshare{s}_b$.}
		\Output{For $b \in \{0, 1\}$, $P_b$ learns $\ashare{z}{L}_b$ s.t. $z = x$ if $s = 0$, else \hspace*{-0.8mm} $z = -x$.}
		\State \multiline{%
			For $b \in \{0, 1\}$, $P_b$ chooses $r_b \overset{\$}{\leftarrow} \mathbb{Z}_{L}$ and computes $w_b = \ashare{x}{L}_b(1 - 2 \cdot \bshare{s}_b) + 2 \cdot r_b$.
		}
		\State \multiline{%
			$P_0$ and $P_1$ invoke an instance of ${2 \choose 1}\text{-COT}_\ell$ where $P_0$ is the sender with correlation function $\rho(r_0) = r_0 + \ashare{x}{L}_0 - 2 \cdot \bshare{s}_0 \cdot \ashare{x}{L}_0$ and $P_1$ is the receiver with input $\bshare{s}_1$. Party $P_1$ learns $y_1 = r_0 + \bshare{s}_1(\ashare{x}{L}_0 - 2 \cdot \bshare{s}_0 \cdot \ashare{x}{L}_0)$.}
		\State \multiline{%
			$P_0$ and $P_1$ invoke an instance of ${2 \choose 1}\text{-COT}_\ell$ where $P_1$ is the sender with correlation function $\rho(r_1) = r_1 + \ashare{x}{L}_1 - 2 \cdot \bshare{s}_1 \cdot \ashare{x}{L}_1$ and $P_0$ is the receiver with input $\bshare{s}_0$. Party $P_0$ learns $y_0 = r_1 + \bshare{s}_0(\ashare{x}{L}_1 - 2 \cdot \bshare{s}_1 \cdot \ashare{x}{L}_1)$.}
		\State For $b \in \{0, 1\}$, $P_b$ outputs $\ashare{z}{L}_b =  w_b - 2 \cdot y_b$.
	\end{algorithmic}
\end{algorithm}

\noindent \textit{Communication Complexity and Security.} Our optimized $\mathcal{F}_{\text{ABS\_MUX}}^{\ell}$ functionality requires only two calls to $\binom{2}{1}\text{-COT}_{\ell}$, resulting in a communication cost of $2(\lambda+\ell)$. This is twice as efficient as the naive approach, which requires four calls and incurs $4(\lambda+\ell)$ communication cost~\cite{rathee_sirnn_2021}, and also improves over the multiplexer in~\cite{rathee_cryptflow2_2020}, which costs $4(\lambda+2\ell)$.

We also note that this combination can be further generalized to realize a ``general multiplexer'' that allows selections between any two inputs $(x, y)$. We provide the description in Appendix~\ref{appendix:gen_mux} for independent interest.

\vspace{0.05in}

\noindent\textbf{MSB}. We employ the protocol of Rathee et al.~\cite{rathee_cryptflow2_2020} to realize $\mathcal{F}_{\text{MSB}}^{\ell}$ (Alg.~\ref{alg:msb}). It takes as input $\ell$-bit arithmetic shares of $x$ and returns the most significant bit of $x$. We set $\ashare{x}{L}_b = m_b || x^{\ell-1}_b$, where $m_b$ denotes the most significant bit of $P_b$'s share of $x$, and $x^{\ell-1}_b$ represents the remaining $\ell-1$ bits of their shares. The MSB of $x$, denoted as $\mathrm{msb}$, can then be calculated as $\mathrm{msb} = m_0 \oplus m_1 \oplus c$, where the Boolean shares of $c$ (the carry bit) are computed by invoking an instance of $\mathcal{F}_{\text{Mill}}^{\ell-1}$.

\begin{algorithm}
	\small
	\caption{Most Significant Bit, $\mathcal{F}^{\ell}_{\text{MSB}}$}\label{alg:msb}
	\begin{algorithmic}[1]
		\Input{For $b \in \{0,1\}$, $P_b$ holds $\ashare{x}{L}_b$.}
		\Output{For $b \in \{0, 1\}$, $P_b$ learns $\bshare{\mathrm{msb}}_b$.}
		\State \multiline{%
			For $b \in \{0,1\}$, $P_b$ parses its share of $x$ such that $\ashare{x}{L}_b = m_b || x^{\ell-1}_b$, where $m_b \in \{0,1\}$.}
		\State \multiline{%
		$P_0$ and $P_1$ invoke an instance of $\mathcal{F}_{\text{Mill}}^{\ell-1}$, where $P_0$'s input is $2^{\ell-1}-1-x_0$ and $P_1$'s input is $x_1$. For $b \in \{0, 1\}$, $P_b$ learns $\bshare{c}_b$.}
		\State \multiline{%
			For $b \in \{0,1\}$, $P_b$ outputs $\bshare{\mathrm{msb}}_b$ = $m_0 \oplus m_1 \oplus c$.}
	\end{algorithmic}
\end{algorithm}

\noindent \textit{Communication Complexity and Security.} In $\mathcal{F}_{\text{MSB}}^{\ell}$, we communicate within the protocol for $\mathcal{F}_{\text{Mill}}^{\ell-1}$, which costs $< (\lambda+14)(\ell-1)$. Security follows directly from that of $\mathcal{F}_{\text{Mill}}^{\ell-1}$ (see Section~\ref{subsec:primitives}).

\subsection{Secure Maximum} \label{subsec:max}

The core cryptographic tool for securely computing \mbox{$L^{\infty}$-norm} (see Section~\ref{sec:l-infinity}) is a protocol to compute shares of the maximum element in an $n$-dimensional vector $x$. We denote this functionality by $\func{Max}{\ell}$.

\subsubsection{Secure Maximum for 2 Elements}
Computing the maximum of $n$ elements requires computing the maximum of a pair $(x_i, x_j)$. To do this, the parties jointly compute $s = 1\{d > 0\}$, where $d = x_i - x_j$, and return $x_i$ if $s = 1$, and $x_j$ otherwise. To obtain Boolean shares of $s$, the parties compute the MSB of $d$ via a call to $\func{MSB}{\ell}$, and then one party complements its share. This ensures that the parties jointly hold correct shares of $s$. They then invoke the $\func{Base\_MUX}{\ell}$ functionality on $\ashare{d}{L}$ and $\bshare{s}$. After this, each party jointly holds shares of $z = d = x_i - x_j$ or $z = 0$, depending on $s$. Finally, each party locally computes $z + x_j$, yielding $\max = x_i$ if $x_i > x_j$, or $\max = x_j$ otherwise.

\subsubsection{Secure Maximum for $n > 2$ Elements}
A naive approach to implement $\func{Max}{\ell}$ for $n > 2$ is to sequentially compare each pair of elements in the vector and update the maximum when necessary. This approach is used in \cite{rathee_cryptflow2_2020} for the MaxPool protocol. Initially, the maximum is set to the first element of the vector $x$, and then, for each $i \in [2..n]$, the maximum is updated to $x[i]$ if $x[i] > \text{max}$. The communication and computation complexity of $\func{Max}{\ell}$ is $O(n)$ due to the $O(n)$ comparisons, but the round complexity is also $O(n)$.

Instead, we employ a binary-tree-like approach that reduces the round complexity of the naive method. The elements of the vector $x$ are organized as leaf nodes of a binary tree. Starting from the leaves, we iteratively compare each pair of elements at the same level and assign their parent node the maximum value. In this way, the root node holds the maximum of $x$. Since comparisons at each level can be done \emph{in batch}, the round complexity reduces from $O(n)$ to the tree depth, $O(\log(n))$. We formally define this functionality as $\mathcal{F}^{\ell}_{\text{Max}}$ (Alg.~\ref{alg:max}).

\begin{algorithm}
	\small
	\caption{Maximum, $\mathcal{F}^{\ell}_{\text{Max}}$}\label{alg:max}
	\begin{algorithmic}[1]
		\Input{For $b \in \{0, 1\}$, $P_b$ holds $\ashare{x}{L}_b \in \mathbb{Z}^{n}_L$.}
		\Output{For $b \in \{0, 1\}$, $P_b$ learns $\ashare{m}{L}_b$, where $m = \underset{i}{\max}{(x_i)}$.}
		\State \multiline{%
			For $b \in \{0,1\}$, party $P_b$ sets $\text{tmp}_{i,b} = \ashare{x_i}{L}_b$, $\forall i \in \{0,\dots,n-1\}$.
		}
		\While{$\text{width} < 1$} \Comment{width = \#nodes at current level}
		\For{$i=0$ to $\text{width}'$} \Comment{width$'$ = \#nodes at next level}
		\If{$\text{width} =_2 0$} \Comment{if \#nodes is even}
		\State \multiline{%
			For $b \in \{0,1\}$, $P_b$ sets $d_{i,b} = \text{tmp}_{2i,b} - \text{tmp}_{2i+1,b}$.
		}
		\Else
		\State \multiline{%
			For $b \in \{0,1\}$, $P_b$ sets $d_{i,b} = \text{tmp}_{2i,b} - \text{tmp}_{2i+1,b}$ if $i < \text{width}' - 1$, and $d_{i,b} = \text{tmp}_{2i,b}$ otherwise.}
		\EndIf
		\State \multiline{%
		$P_0$ and $P_1$ invoke an instance of $\mathcal{F}^{\ell}_{\text{MSB}}$. For $b \in \{0,1\}$, $P_b$ inputs $d_{i,b}$ and learns $\bshare{\mathrm{msb}_i}_b$.}
		\State \multiline{%
			$P_0$ sets $\bshare{s_i}_0 = \bshare{\mathrm{msb}_i}_0$, and $P_1$ sets $\bshare{s_i}_1 = \bshare{\mathrm{msb}_i}_1 \oplus 1$. Note that $s_i = \ov{\text{msb}$_{i}$}$.}
		\State \multiline{%
			$P_0$ and $P_1$ invoke an instance of $\func{Base\_MUX}{\ell}$ with inputs $(d_{i,0},\bshare{s_i}_0)$ and ($d_{i,1},\bshare{s_i}_1)$, respectively. For $b \in \{0,1\}$, $P_b$ learns $\ashare{z_i}{L}_b$, where $z_i = s_i \cdot d_i$ and $d_i = d_{i,0} + d_{i,1}$.}
		\If{$\text{width} =_2 0$}
		\State \multiline{%
			For $b \in \{0,1\}$, $P_b$ sets $\text{tmp}_{i,b} = \ashare{z_i}{L}_b + \text{tmp}_{2i+1,b}$.
		}
		\Else
		\State \multiline{%
			For $b \in \{0,1\}$, $P_b$ sets $\text{tmp}_{i,b} = \ashare{z_i}{L}_b + \text{tmp}_{2i+1,b}$ if $i < \text{width}' - 1$; otherwise, $\text{tmp}_{i,b} = \ashare{z_i}{L}_b$.
		}
		\EndIf
		\EndFor
		\EndWhile
		\State For $b \in \{0,1\}$, $P_b$ outputs $\ashare{m}{L}_b$, where $m = \text{tmp}_{0,0} + \text{tmp}_{0,1}$.
	\end{algorithmic}
\end{algorithm}

\vspace{0.05in}

\noindent\textit{Communication Complexity and Security.} Alg.~\ref{alg:max} makes \(\sum_{i=1}^{\lceil \log n \rceil} \lfloor n_i/2 \rfloor = n-1\) calls each to \(\func{MSB}{\ell}\) and \(\func{Base\_MUX}{\ell}\), where \(n_i\) is the number of nodes at level \(i\). These incur costs of \((n-1)(2\lambda + 2\ell)\) and \(< (n-1)(\lambda + 14)(\ell - 1)\), respectively, for a total of \(< (n-1)(\lambda + 16)(\ell + 1) - 30(n-1)\). Security directly follows from that of \(\func{MSB}{\ell}\) and \(\func{Base\_MUX}{\ell}\).

\begin{table*}[htbp]
\centering
\small
\renewcommand{\arraystretch}{1.3}
\caption{Summary of our 2PC building blocks. All communication is in bits.}
\small
\begin{tabular}{|>{\centering\arraybackslash}m{2.8cm}|c|c|>{\centering\arraybackslash}m{2.5cm}|>{\centering\arraybackslash}m{3.1cm}|}
\hline
Functionality & Notation & Objective & Description & Comm. Complex. \\ \hline
\addlinespace[1.5pt] \hline
Most Significant Bit & $\ashare{\text{msb}}{L}_b = \func{MSB}{\ell}(\ashare{x}{L}_b)$ & \texttt{msb = MSB(x)}  & Returns MSB of $x$ & $< (\lambda+14)(\ell-1)$ \\ \hline
Multiplexer for ABS & $\ashare{z}{L}_b = \func{ABS\_MUX}{\ell}(\bshare{s}_b, \ashare{x}{L}_b)$ & \texttt{s ? -x : x} & Returns $-x$ if $s$ is 1 and $x$ otherwise & $2\lambda + 2\ell$ \\ \hline
Absolute Value & $\ashare{\text{abs}}{L}_b = \func{ABS}{\ell}(\ashare{x}{L}_b)$ & \texttt{(x > 0) ? x : -x} & Returns $x$ if $x>0$ and $-x$ otherwise & $< (\lambda+16)(\ell+1) - 30$ \\ \hline
Base Multiplexer & $\ashare{z}{L}_b = \func{Base\_MUX}{\ell}(\bshare{s}_b, \ashare{x}{L}_b)$ & \texttt{s ? x : 0} & Returns $x$ if $s$ is $1$ and $0$ otherwise & $2\lambda + 2\ell$ \\ \hline
Maximum  & $\ashare{m}{L}_b = \func{Max}{\ell}(\ashare{x}{L}_b), x \in \mathbb{Z}^n_L$ & \texttt{max(x)} & Returns $\underset{i}{\max}{(x_i)}$ & $< (n-1)(\lambda+16)(\ell+1)-30(n-1)$ \\ \hline
\end{tabular}
\label{tab:algs-summary}
\end{table*}

\section{Protocol Design for $L^p$-norms} \label{sec:norm_protocols}

We are now ready to present our protocol constructions for common $L^p$-norms: $L^1$, $L^2$, and $L^\infty$. These norms are widely used as distance metrics in applications such as secure machine learning (e.g., clustering, inference), biometric matching, and location-based services (see Sections~\ref{sec:intro} and~\ref{sec:related}).

\subsection{Protocol for $L^1$-Norm} \label{subsec:l1_norm}

The $L^1$-norm (also known as the Manhattan distance) measures the distance between two points along axis-aligned paths. Formally, given a vector $x = (x_1, x_2, \dots, x_n)$, the $L^1$-norm is defined as $||x||_1 = \sum_{i=1}^{n} |x_i|$. This norm represents the distance from $x$ to the origin when moving along grid-based paths. More generally, we consider distance from an arbitrary point $y = (y_1, \dots, y_n)$, defined as $||x - y||_1 = \sum_{i=1}^{n} |x_i - y_i|$. Alg.~\ref{alg:md} realizes the secure functionality $\mathcal{F}^{\ell}_{\text{MD}}$, allowing two parties to jointly compute $||x - y||_1$ from secret shares of $n$-dimensional vectors $x$ and $y$, whose entries are $\ell$-bit integers.

\begin{algorithm}
\small
\caption{Manhattan Distance ($L^1$-Norm), $\mathcal{F}^{\ell}_{\text{MD}}$}\label{alg:md}
\begin{algorithmic}[1]
\Input{For $b \in \{0, 1\}$, $P_b$ holds $\ashare{x}{L}_b \in \mathbb{Z}^{n}_L$ and $\ashare{y}{L}_b \in \mathbb{Z}^{n}_L$.}
\Output{For $b \in \{0, 1\}$, $P_b$ learns $\ashare{\sum\limits_{i=1}^{n}|x_i-y_i|}{L}_b$.}
\For{$i=0$ to $n-1$}
\State \multiline{%
For $b \in \{0,1\}$, $P_b$ sets $\ashare{d_i}{L}_b = \ashare{x_i}{L}_b - \ashare{y_i}{L}_b$, where $d_i = x_i - y_i$.}
\State \multiline{%
$P_0$ and $P_1$ invoke an instance of $\mathcal{F}^{\ell}_{\text{ABS}}$, with inputs $\ashare{d_i}{L}_0$ and $\ashare{d_i}{L}_1$, respectively. For $b \in \{0, 1\}$, $P_b$ obtains $\ashare{\mathrm{abs}_i}{L}_b$, where $\mathrm{abs}_i = |d_i|$.}
\EndFor
\State For $b \in \{0, 1\}$, $P_b$ outputs $\sum\limits_{i=1}^{n}\ashare{\mathrm{abs}}{L}_b = \ashare{\sum\limits_{i=1}^{n}\mathrm{abs}}{L}_b$.
\end{algorithmic}
\end{algorithm}


\subsection{Protocol for $L^2$-Norm} \label{sec:l2}
The $L^2$-norm (also known as the Euclidean norm) measures the shortest distance to the origin and is the most commonly used norm. It is defined as the square root of the sum of squares of a vector’s components: $||x||_2 = \sqrt{\sum_{i=1}^{n} x_i^2}$, where $x$ is an $n$-dimensional vector. We consider the general case where the origin is replaced with another vector $y$, yielding the Euclidean distance: $||x - y||_2 = \sqrt{\sum_{i=1}^{n} (x_i - y_i)^2}$. Alg.~\ref{alg:ed} realizes the functionality $\func{ED}{\ell}$ to securely compute $||x - y||_2$, where the parties hold shares of $\ell$-bit integers $x_i$ and $y_i$ for $i \in \{1, \dots, n\}$. At the end of the protocol, each party $P_b$ holds a share of $||x - y||_2$. Alg.~\ref{alg:ed} uses the multiplication and square root protocols from~\cite{rathee_sirnn_2021}, denoted as $\func{Mult}{\ell}$ and $\func{SQRT}{\ell}$, respectively.

\begin{algorithm}
\small
\caption{Euclidean Distance ($L^2$-Norm), $\mathcal{F}^{\ell}_{\text{ED}}$}\label{alg:ed}
\begin{algorithmic}[1]
\Input{For $b \in \{0, 1\}$, $P_b$ holds $\ashare{x}{L}_b \in \mathbb{Z}^{n}_L$ and $\ashare{y}{L}_b \in \mathbb{Z}^{n}_L$.}
\Output{For $b \in \{0, 1\}$, $P_b$ learns $\ashare{z}{L}_b$, and $z = \sqrt{\sum\limits_{i=1}^{n} (x_i-y_i)^2}$.}
\For{$i=0$ to $n-1$}
\State \multiline{%
For $b \in \{0,1\}$, $P_b$ sets $\ashare{d_i}{L}_b = \ashare{x_i}{L}_b - \ashare{y_i}{L}_b$, where $d_i = x_i - y_i$.}
\State \multiline{%
$P_0$ and $P_1$ invoke an instance of \vspace{1pt} $\func{Mult}{\ell}$ to compute shares of $d^2_i$. For $b \in \{0,1\}$, $P_b$ learns $\ashare{d^2_i}{L}_b$.}
\EndFor
\State For $b \in \{0,1\}$, $P_b$ computes $\ashare{d^2}{L}_b = \sum\limits_{i=1}^{n} \ashare{d^2_i}{L}_b$.
\State \multiline{%
$P_0$ and $P_1$ invoke an instance of $\func{SQRT}{\ell}$ to compute shares of $\ashare{d}{L}_b = \sqrt{\ashare{d^2}{L}_b}$.} \label{step:sqrt}
\State For $b \in \{0,1\}$, $P_b$ outputs $\ashare{d}{L}_b$.
\end{algorithmic}
\end{algorithm}


\subsection{Protocol for $L^\infty$-Norm}
\label{sec:l-infinity}
The $L^\infty$-norm (also known as the max norm or Chebyshev distance) is defined as the absolute value of the largest component of a vector. For an $n$-dimensional vector $x$, we have: $||x||_\infty = \underset{i}{\max}{(|x_i|)}$. Given an $n$-dimensional origin $y$, the Chebyshev distance measures the largest magnitude among the components of the difference vector: $||x - y||_\infty = \underset{i}{\max}{(|x_i - y_i|)}$. In our setting, parties $P_0$ and $P_1$ hold arithmetic secret shares of $x$ and $y$ and aim to jointly compute shares of the maximum absolute difference between corresponding elements. To realize this functionality, denoted by $\func{CD}{\ell}$, the parties first locally compute shares of the difference vector $d = x - y$, which incurs no communication. They then invoke $\func{ABS}{\ell}$ on the shares of $d$, followed by a call to the secure maximum functionality $\func{Max}{\ell}$. Alg.~\ref{alg:cd} outlines the protocol.

\begin{algorithm}
\small
\caption{Chebyshev Distance $(L^\infty$-Norm), $\mathcal{F}^{\ell}_{\text{CD}}$}\label{alg:cd}
\begin{algorithmic}[1]
\Input{For $b \in \{0, 1\}$, $P_b$ holds $\ashare{x}{L}_b \in \mathbb{Z}^{n}_L$ and $\ashare{y}{L}_b \in \mathbb{Z}^{n}_L$.}
\Output{For $b \in \{0, 1\}$, $P_b$ gets $\ashare{m}{L}_b$ where $m = \underset{i}{\max}{(|x_i - y_i|)}$.}
\For{$i=0$ to $n-1$}
\State \multiline{%
For $b \in \{0,1\}$, $P_b$ sets $\ashare{d_i}{L}_b = \ashare{x_i}{L}_b - \ashare{y_i}{L}_b$, where $d_i = x_i - y_i$.}
\State \multiline{%
$P_0$ and $P_1$ invoke an instance of $\mathcal{F}^{\ell}_{\text{ABS}}$, with inputs $\ashare{d_i}{L}_0$ and $\ashare{d_i}{L}_1$. For $b \in \{0, 1\}$, $P_b$ obtains $\ashare{d'_i}{L}_b$, where $d'_i = |d_i|$.}
\EndFor
\State \multiline{%
$P_0$ and $P_1$ invoke an instance of $\mathcal{F}^{\ell}_{\text{Max}}$ functionality, with inputs $\ashare{d'}{L}_0$ and $\ashare{d'}{L}_1$. For $b \in \{0, 1\}$, $P_b$ obtains $\ashare{m}{L}_b$, where $m = \underset{i}{\max}{(d'_i)}$.} 
\State For $b \in \{0,1\}$, $P_b$ outputs $\ashare{m}{L}_b$, where $m = \underset{i}{\max}{(|x_i - y_i|)}$.
\end{algorithmic}
\end{algorithm}

\section{\sys for Secure Inferences} \label{sec:secure_addernet}

In this section, we will discuss in detail the novel application of our \sys to secure inferences.\footnote{To our best knowledge, we take the first step to design and integrate secure $L^p$-norm for secure ML inferences, and develop more efficient secure $L^p$-norm for other applications (e.g., clustering, biometric matching).} 

\subsection{$L^p$-Norm-based CNN Inferences}
In CNNs, a convolutional layer uses a filter $F \in \mathbb{R}^{d \times d \times c_{in} \times c_{out}}$, where $d$ is the kernel size, and $c_{in}$ and $c_{out}$ are the numbers of input and output channels, respectively. The input tensor is $X \in \mathbb{R}^{h_{in} \times w_{in} \times c_{in}}$, with height $h_{in}$ and width $w_{in}$. The convolution operation generalizes to computing the similarity between input and filter~\cite{addernet}:

\begin{equation} \label{eq:feature_extraction}
\text{Y}(m,n,t) = \sum_{i=1}^{d}\sum_{j=1}^{d}\sum_{k=1}^{c_{in}} \text{S}(X_{m + i, n + j, k}, F_{i, j, k, t}),
\end{equation}
\noindent where $\text{S}(\cdot, \cdot)$ is a similarity function (e.g., convolution).

Cross-correlation, typically denoted by the \textit{Conv} operation, is the standard similarity metric in CNNs. However, other metrics, such as $L^p$-norms, can also be employed. Chen et al.~\cite{addernet} observe that cross-correlation closely resembles the $L^2$-norm in capturing feature similarity and that both yield comparable accuracy with appropriate adaptation. They further demonstrate that replacing convolution with the $L^1$-norm preserves accuracy while significantly reducing the number of multiplications. This leads to the Adder Neural Network (AdderNet), which adopts a new similarity function called the \textit{Adder} operation, defined as $\text{Adder}(x, y) = -|x - y|$.

\subsection{AdderNet ($L^1$-Norm-based)}
In a series of works \addernet{}, Chen et al.~\cite{addernet} proposed AdderNet, which aims to construct an almost ``multiplication-free'' neural network. This is achieved by replacing convolutional layers, which involve numerous multiplications, with ``\textit{Adder}'' layers that compute similarity between input features and filters using additions and $L^1$-norm operations.

Standard convolutions use cross-correlation (denoted by the \textit{Conv} operation), where $\text{Conv}(x, y) = x \cdot y$ for $\ell$-bit integers $x$ and $y$. In contrast, AdderNet uses the negative $L^1$-norm (denoted by the \textit{Adder} operation), where $\text{Adder}(x, y) = -|x - y|$. Accordingly, Eq.~\ref{eq:feature_extraction} is modified as:

\begin{align} \label{eq:adder}
&\text{Y}(m,n,t) = \sum_{i=1}^{d}\sum_{j=1}^{d}\sum_{k=1}^{c_{in}} \text{Adder}(X_{m + i, n + j, k}, F_{i, j, k, t}) \nonumber\\
&\phantom{{}\text{Y}(m,n,t){}} = \sum_{i=1}^{d}\sum_{j=1}^{d}\sum_{k=1}^{c_{in}} -|X_{m + i, n + j, k} - F_{i, j, k, t}|
\end{align}

Since the \textit{Adder} output is always negative, batch normalization is applied, introducing minimal multiplications with negligible performance impact~\cite{addernet}. While the \textit{Conv} operation requires $O(h_{out}w_{out}d^2c_{in}c_{out})$ multiplications, the \textit{Adder} operation reduces this to $O(h_{out}w_{out}c_{out})$, limited to batch normalization. Thus, \textit{Adder} avoids $d^2c_{in}$ multiplications per output element, significantly reducing computational complexity.

\subsection{Secure 2PC Inference} Next, we build our secure protocol for the \textit{Adder} operation based on the $L^1$-norm protocol from Section~\ref{sec:norm_protocols}. To this end, we define a functionality $\func{Adder}{\ell}$ that, given inputs $x, y \in \mathbb{Z}_L$ shared between parties $P_0$ and $P_1$, outputs shares of the negative Manhattan distance between $x$ and $y$. Alg.~\ref{alg:adder} realizes this functionality.

\begin{algorithm}
\small
\caption{Adder Operation, $\mathcal{F}^{\ell}_{\text{Adder}}$}\label{alg:adder}
\begin{algorithmic}[1]
\Input{For $b \in \{0, 1\}$, party $P_b$ holds $\ashare{x}{L}_b \in \mathbb{Z}^{1}_L$ and $\ashare{y}{L}_b \in \mathbb{Z}^{1}_L$.}
\Output{For $b \in \{0, 1\}$, party $P_b$ learns $\ashare{-|x-y|}{L}_b$.}
\State \multiline{%
$P_0$ and $P_1$ invoke an instance of $\func{MD}{\ell}(\ashare{x}{L}, \ashare{y}{L})$ and learn shares of $d = |x - y|$.
}
\State \multiline{%
For $b \in \{0, 1\}$, $P_b$ sets $\ashare{d'}{L}_b = -\ashare{d}{L}_b$.
}
\State For $b \in \{0, 1\}$, $P_b$ outputs $\ashare{d'}{L}_b$, where $d' = -|x - y|$.
\end{algorithmic}
\end{algorithm}
\section{Evaluation} \label{sec:evals}



\subsection{Experimental Setup}

All experiments ran on a server with an AMD Threadripper PRO (64 threads) and 512GB RAM. We emulated hosts/nodes for different parties in a LAN with 51 Gbps bandwidth and ~0.033 ms round-trip latency. The system runs Ubuntu 22.04.1 LTS, with all programs compiled using GCC 9.5.0.

\vspace{0.05in}

\noindent\textbf{Datasets}.
We evaluate \sys across a range of applications using the following datasets. For general benchmarking, we generate synthetic data with varying dimensions. For biometric matching, we use the ``AT\&T Database of Faces''~\cite{samaria_parameterisation_1994}, containing 400 images of size $92 \times 112$. In the context of location-based services, we employ a single-building indoor localization dataset~\cite{montoliu_indoorloc_2017, yang_death_2018, jarvinen_pilot_2019} with 505 reference points and 241 access points, and a large-scale spatial crowdsourcing dataset from Twitter~\cite{yuan_priradar_2020} with 1 million requests and 2 million workers. For secure ML tasks, we use the Lsun dataset~\cite{ultsch_fundamental_2020} for clustering (400 vectors), and the SIFT~\cite{lowe_object_1999}, Deep1B~\cite{yandex_efficient_2016}, and Amazon~\cite{mcauley_image-based_2015} datasets for secure $k$-NN. Finally, we evaluate secure inference using the CIFAR-10 dataset~\cite{Krizhevsky2009LearningML}.

\subsection{Overview for Crypto-$L^p$}

\begin{table*}[htbp]
\caption{Summary of our benchmarks and the $L^p$-distance metrics used in them. ($\ast$) indicates that we designed our own $L^p$-norms.}
\scriptsize
\centering
\setlength{\tabcolsep}{3pt}
\renewcommand{\arraystretch}{0.4}
\begin{NiceTabular}{@{}|c|>{\centering\arraybackslash}m{3cm}|c|c|>{\centering\arraybackslash}m{3cm}|@{}}
\toprule
Context & Application(s) & Systems & $L^p$-Distance Metric(s) & Cryptographic Scheme \\ \midrule
\multirow{2}{*}{\parbox[c]{3cm}{\centering Secure Location-based Services}} & Indoor Localization & PILOT \cite{jarvinen_pilot_2019} & $L^1$ and $L^2$ & Mixed (OT\hspace{1pt}+\hspace{1pt}GC\hspace{1pt}+\hspace{1pt}GMW) \\ \cmidrule{2-5}
& Spatial Crowdsourcing & \etal{Han} \cite{han_location_2020} & $L^1$, $L^2$ and $L^\infty$ & HE-based \\ \cmidrule{1-5}
\multirow{2}{*}{\parbox[c]{3cm}{\centering Secure \\ Pattern Recognition}} & \multirow{3}{*}{Biometric Matching} & ABY \cite{demmler_aby_2015} & $L^2$ & Mixed (OT\hspace{1pt}+\hspace{1pt}GC) \\ \cmidrule{3-5}
& & GSHADE \cite{bringer_gshade_2014} & $L^2$ & OT-based \\ \cmidrule{1-5}
\multirow{18}{*}{\parbox[c]{3cm}{\centering Secure \\ Machine Learning}} & \multirow{3}{*}{K-means Clustering} & \etal{Mohassel} \cite{mohassel_practical_2019} & $L^1$, $L^2$ and $L^\infty$ & Mixed (OT\hspace{1pt}+\hspace{1pt}GC) \\ \cmidrule{3-5}
& & \etal{Jäschke} \cite{jaschke_unsupervised_2019} & $L^1$ & HE-based \\ \cmidrule{2-5}
& \multirow{3}{*}{$k$-Nearest Neighbors} & \etal{Cong} \cite{cong_revisiting_2023} & $L^2$ & HE-based \\ \cmidrule{3-5}
& & SANNS \cite{chen_sanns_2020} & $L^2$ & HE-based \\ \cmidrule{2-5}
& \multirow{3}{*}{\parbox[c]{3cm}{\centering Convolutional Neural Networks}} & Cheetah \cite{huang_cheetah_2022}, CrypTFlow2 \cite{rathee_cryptflow2_2020} & $\ast$ & Mixed (OT\hspace{1pt}+\hspace{1pt}HE) \\ \cmidrule{3-5}
& & Delphi \cite{mishra_delphi_2020}, Gazelle \cite{juvekar_gazelle_2018} & $\ast$ & Mixed (HE\hspace{1pt}+\hspace{1pt}GC) \\ \cmidrule{2-5}
& Neural Networks, Regression & ABY3 \cite{mohassel_aby3_2018}& $\ast$ & Mixed (OT\hspace{1pt}+\hspace{1pt}GC) \\ \bottomrule
\end{NiceTabular}
\label{tab:benchmarks_summary}
\end{table*}

Similar to SOTA cryptographic systems, we utilize primitives such as OT extensions, the millionaire's problem, and secure square root (SQRT) from the SCI library~\cite{rathee_cryptflow2_2020} \footnote{\url{https://github.com/mpc-msri/EzPC/tree/master/SCI}} to build our cryptographic building blocks and systems. Moreover, we optimize our shared building blocks within SCI, such as the maximum protocol, and report the performance comparison. Although the SCI library and related extensions \cite{huang_cheetah_2022, rathee_sirnn_2021, rathee_cryptflow2_2020, rathee_secfloat_2022} are primarily designed for secure inference and not directly relevant to our benchmarks, we leverage these optimized primitives due to their superior performance over prior works \cite{mishra_delphi_2020, juvekar_gazelle_2018, mohassel_secureml_2017, demmler_aby_2015}. 

In Table~\ref{tab:microbenchmarks}, we report the performance of Crypto-$L^p$ and its building blocks in runtime and communication for input vectors of dimension $2^{16}$. As discussed in Section~\ref{sec:l2}, we employ the SQRT protocol from \cite{rathee_sirnn_2021} for our $L^2$-norm. For the remainder of this section, $L^2$-norm or Euclidean distance will specifically refer to the squared $L^2$-norm.

\begin{table}[!h]
\small
\centering
\setlength{\tabcolsep}{4pt}
\caption{\sys and its building blocks on input vectors of dimension $2^{16}$, consisting of $32$-bit integers.}
\small
\begin{NiceTabular}{@{}cccccc@{}}
\toprule
\multirow{2}{*}{Norm} & \multirow{2}{*}{Building Blocks} & \multicolumn{2}{c}{Runtime (ms)} & \multicolumn{2}{c}{Comm. (MB)} \\ \cmidrule(lr){3-4} \cmidrule(lr){5-6}
&  & Block & Total & Block & Total \\ \midrule
\multirow{2}{*}{$L^1$} & MSB & $318$ & \multirow{2}{*}{$324$} & $29.97$ & \multirow{2}{*}{$32.47$} \\
& ABS MUX & $6$ &  & $2.50$ & \\ \midrule
\multirow{2}{*}{$L^2$} & Mult & $234$ & \multirow{2}{*}{$6135$} & $74$ & \multirow{2}{*}{$747.94$} \\
& SQRT & $5901$ &  & $673.94$ & \\ \midrule
Squared $L^2$ & Mult & $234$ & $234$ & $74$ & $74$ \\ \midrule
\multirow{2}{*}{$L^\infty$} & ABS & $324$ & \multirow{2}{*}{$688$} & $29.97$ & \multirow{2}{*}{$63.23$} \\
& Max & $364$ &  & $33.26$ & \\ \bottomrule
\end{NiceTabular}
\vspace{-0.15in}
\label{tab:microbenchmarks}
\end{table}

\begin{figure}[!h]
\small
\centering
\subfigure[Runtime \label{fig:max_opti_vs_naive_runtime}]{
\includegraphics[width=0.4625\columnwidth]{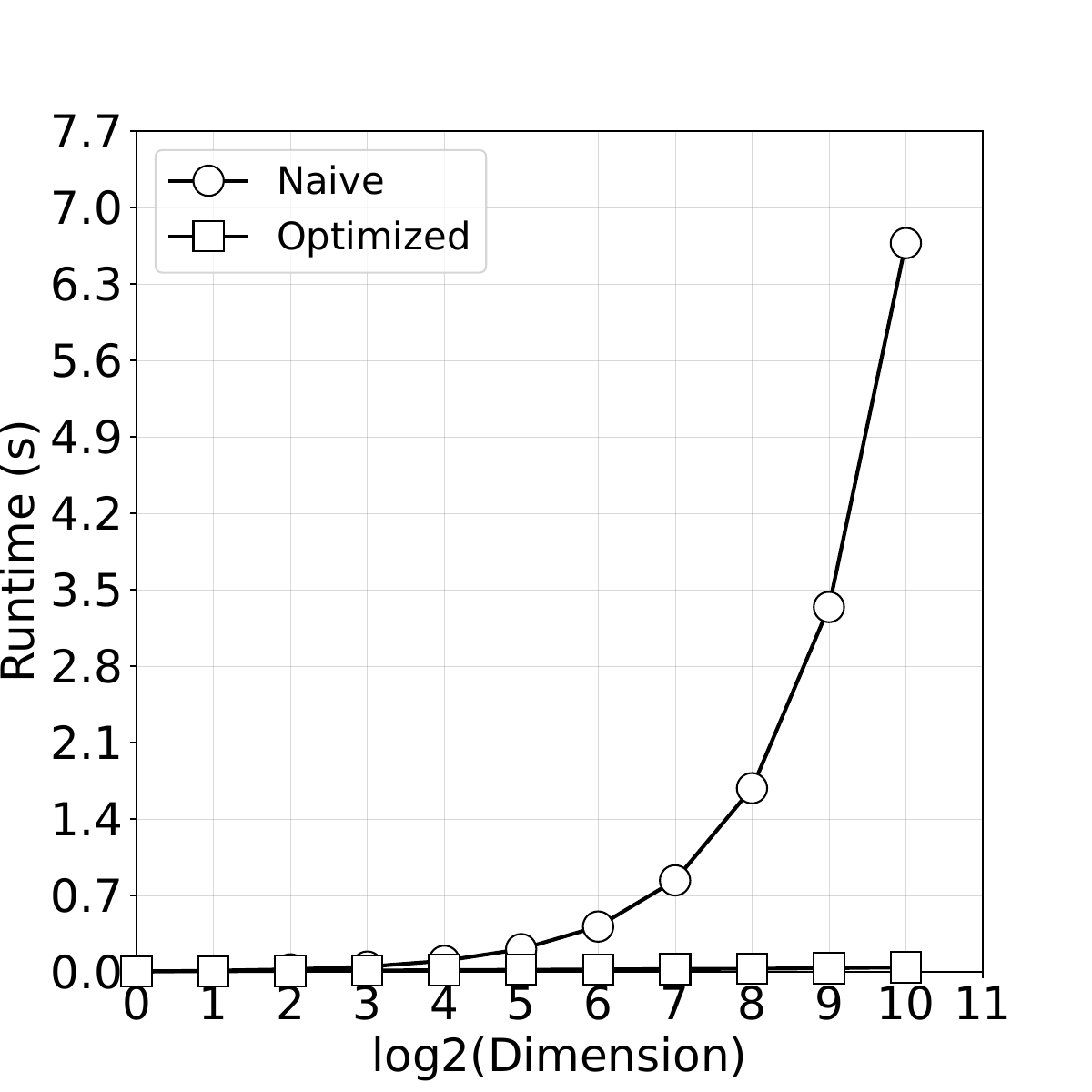}}
\subfigure[Communication \label{fig:max_opti_vs_naive_comm}]{
\includegraphics[width=0.4625\columnwidth]{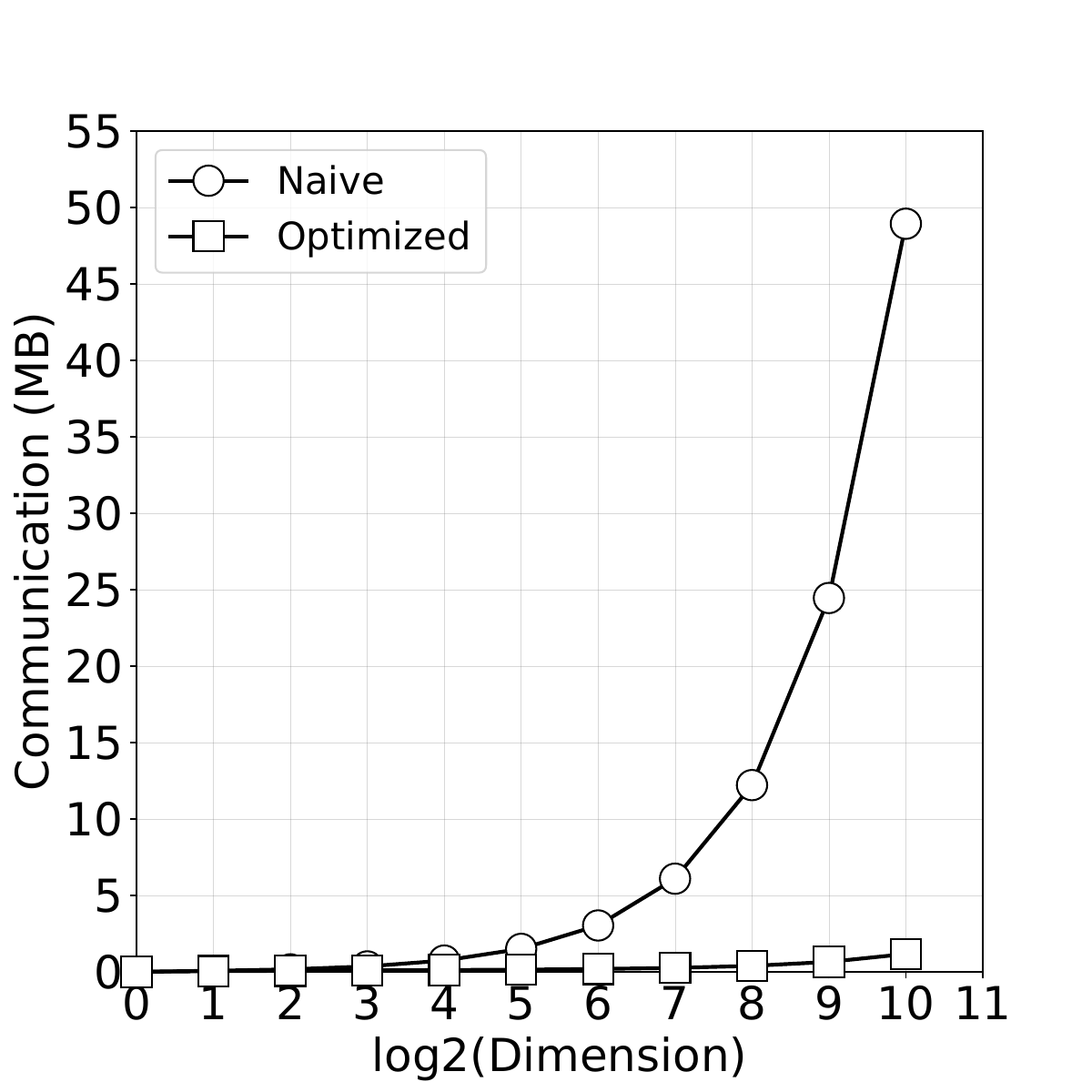}}
\caption{Runtime and communication improvements of our optimized maximum protocol over the naive protocol (as in \cite{rathee_cryptflow2_2020}), with a logarithmic x-axis for input size.}
\label{fig:max_opti_vs_naive}
\end{figure}

Figure~\ref{fig:max_opti_vs_naive} compares our optimized maximum protocol with the naive version (similar to the \emph{maxpool} protocol in~\cite{rathee_cryptflow2_2020}) in terms of communication and runtime. The input vector consists of 32-bit integers across dimensions from $1$ to $2^{10}$. Our protocol, which uses a binary tree structure, outperforms the naive approach in both metrics, as shown in Figures~\ref{fig:max_opti_vs_naive_runtime} and~\ref{fig:max_opti_vs_naive_comm}. The runtime improvement stems from reduced round complexity via the tree-based design. For communication, we replace OT with COT and pack 8 selection bits into a single byte by processing pairwise comparisons in parallel at each tree level. In contrast, the naive protocol communicates one byte per bit. This optimization results in significant communication savings.

We next present the performance improvements of \sys{} over SOTA works~\cite{han_location_2020, mohassel_practical_2019, chen_sanns_2020, cong_revisiting_2023, emp-toolkit, jarvinen_pilot_2019, demmler_aby_2015} across Crypto-$L^1$, Crypto-$L^2$, and Crypto-$L^\infty$ in their respective applications. Mohassel et al.~\cite{mohassel_practical_2019} implemented all three norms for secure clustering, making their work the most directly comparable to ours. We also include additional SOTA baselines that use different norms in other applications and compare them with \sys{} in the relevant contexts. Finally, we evaluate our norm protocols using primitives from various inference libraries and toolkits to provide a comprehensive benchmark. Table~\ref{tab:benchmarks_summary} summarizes these benchmarks.

\subsection{Evaluations for Crypto-$L^1$}
We benchmark the communication and runtime of Crypto-$L^1$ against SOTA works using synthetic vectors of size $2^{16} \times$ 32-bit integers for fair comparison. We also evaluate Crypto-$L^1$ on real-world datasets and compare it with prior benchmarks in their respective application contexts. See Table~\ref{tab:merged_l1_comparison_final}.

\begin{table}[!ht]
\setlength{\tabcolsep}{2pt}
\small
\centering
\caption{Comparison of Crypto-$L^1$ and prior works across various secure computation tasks.}
\begin{NiceTabular}{@{}c c c c c c@{}}
\toprule
Systems & Protocol & Runtime (ms) & Speedup & Comm. (MB) & Improv. \\
\midrule
\multicolumn{6}{l}{\textbf{Secure $L^1$ Norm (Synthetic $2^{16} \times 32$ Dataset)}} \\
\cite{demmler_aby_2015} & $\text{2PC}_\text{GC}$ & \(5.90 \times 10^4\) & $182\times$ & \(2.96 \times 10^2\) & $9\times$ \\
\cite{han_location_2020} & $\text{SMPC}_\text{HE}$ & \(6.60 \times 10^7\) & $10^5\times$ & \(6.40 \times 10^2\) & $19\times$ \\
\cite{mohassel_practical_2019} & $\text{2PC}_\text{GC}$ & \(2.67 \times 10^4\) & $82\times$ & \(1.19 \times 10^3\) & $36\times$ \\
\cite{demmler_aby_2015} & $\text{2PC}_\text{GMW}$ & \(9.48 \times 10^4\) & $292\times$ & \(5.32 \times 10^2\) & $16\times$ \\
\cite{mishra_delphi_2020} & $\text{2PC}_\text{GC}$ & \(7.43 \times 10^3\) & $22\times$ & \(6.55 \times 10^3\) & $201\times$ \\
Crypto-$L^1$ & $\text{2PC}_\text{OT}$ & \(3.24 \times 10^2\) & - & \(3.25 \times 10^1\) & - \\
\midrule
\multicolumn{6}{l}{\textbf{Secure Clustering (LSun Dataset)}} \\
\cite{mohassel_practical_2019} & $\text{2PC}_\text{GC}$ & \(2.61 \times 10^2\) & $33\times$ & \(7.00 \times 10^0\) & $17\times$ \\
Crypto-$L^1$ & $\text{2PC}_\text{OT}$ & \(8.00 \times 10^0\) & - & \(4.20 \times 10^{-1}\) & - \\
\midrule
\multicolumn{6}{l}{\textbf{Indoor LBS (Single-Building Dataset)}} \\
\cite{jarvinen_pilot_2019} & $\text{2PC}_\text{GC}$ & \(1.42 \times 10^5\) & $224\times$ & \(5.50 \times 10^2\) & $9\times$ \\
\cite{jarvinen_pilot_2019} & $\text{2PC}_\text{GMW}$ & \(1.69 \times 10^5\) & $267\times$ & \(9.87 \times 10^2\) & $16\times$ \\
Crypto-$L^1$ & $\text{2PC}_\text{OT}$ & \(6.30 \times 10^2\) & - & \(6.23 \times 10^1\) & - \\
\midrule
\multicolumn{6}{l}{\textbf{Spatio-Crowdsourcing (Twitter Dataset)}} \\
\cite{han_location_2020} & $\text{SMPC}_\text{HE}$ & \(6.98 \times 10^7\) & $6348\times$ & \(4.28 \times 10^4\) & $40\times$ \\
Crypto-$L^1$ & $\text{2PC}_\text{OT}$ & \(1.08 \times 10^4\) & - & \(1.06 \times 10^3\) & - \\
\bottomrule
\end{NiceTabular}
\label{tab:merged_l1_comparison_final}
\end{table}

\noindent\textbf{Secure Clustering.}
For $L^1$-based secure clustering, we benchmark against Mohassel et al.~\cite{mohassel_practical_2019}, the leading work in this domain (noting that \cite{jaschke_unsupervised_2019} is not open-sourced). We compare Crypto-$L^1$ with their secure $L^1$-norm implementation under identical settings ($32$-bit integers, $2^{16}$ dimensions). Crypto-$L^1$ achieves up to $82\times$ faster runtime and $36\times$ lower communication. We also compare both systems on secure clustering over the Lsun dataset~\cite{ultsch_fundamental_2020}.

\vspace{0.05in}

\noindent\textbf{Secure LBS.}
We evaluate Crypto-$L^1$ against PILOT~\cite{jarvinen_pilot_2019} and Han et al.~\cite{han_location_2020}, two leading LBS baselines. PILOT uses ABY~\cite{demmler_aby_2015} to implement Manhattan distance via Yao's and GMW sharing. We re-implemented their protocols using ABY, and Crypto-$L^1$ achieves up to $182\times$ faster runtime and $9\times$ lower communication than their GC-based version, and $292\times$ and $16\times$ improvements over their GMW-based version, respectively. For Han et al., whose protocol assumes a multi-party setting with a trusted third party, we emulate a 2PC setting and derive performance using their reported complexity (Tables III–V in~\cite{han_location_2020}), under the same parameters (128-bit security and vector size $2^{16}$). Crypto-$L^1$ reduces communication by up to $19\times$. We further compare Crypto-$L^1$ with PILOT on indoor localization using the widely used single-building database~\cite{montoliu_indoorloc_2017, yang_death_2018, jarvinen_pilot_2019} (505 reference points, 241 access points), and with Han et al.'s spatial crowdsourcing protocol~\cite{han_location_2020} using the Twitter-based dataset from~\cite{yuan_priradar_2020} (1M user requests, 2M workers).

\vspace{0.05in}

\noindent\textbf{Other Benchmarks.}
Although secure inference libraries are not tailored for norm computation, we re-implemented Crypto-$L^1$ using garbled circuits in ABY3~\cite{mohassel_aby3_2018} and Delphi~\cite{mishra_delphi_2020}, both of which use GC for non-linear layers like ReLU. Crypto-$L^1$ outperforms both baselines by up to $22\times$ in runtime and $201\times$ in communication.

\subsection{Evaluations for Crypto-$L^2$}
Similarly, we benchmark the communication cost and runtime of Crypto-$L^2$ with SOTA works (see Table~\ref{tab:merged_l2_all_final}).

\begin{table}[!ht]
\setlength{\tabcolsep}{2pt}
\scriptsize
\centering
\caption{Comparison of Crypto-$L^2$ and prior works across various secure computation tasks.}
\begin{NiceTabular}{@{}c c c c c c@{}}
\toprule
Systems & Protocol & Runtime (ms) & Speedup & Comm. (MB) & Improv. \\
\midrule
\multicolumn{6}{l}{\textbf{Secure $L^2$ Norm (Synthetic $2^{16} \times 32$ Dataset)}} \\
\cite{demmler_aby_2015,jarvinen_pilot_2019} & $\text{2PC}_\text{GC}$ & \(3.98 \times 10^{4}\) & $170\times$ & \(6.30 \times 10^{3}\) & $26\times$ \\
\cite{han_location_2020} & $\text{SMPC}_\text{HE}$ & \(4.50 \times 10^{7}\) & $10^5 \times \uparrow$ & \(2.66 \times 10^{4}\) & $0.35 \times \downarrow$ \\
\cite{mohassel_practical_2019} & $\text{2PC}_\text{OT}$ & \(6.35 \times 10^{4}\) & $271\times$ & \(3.06 \times 10^{2}\) & $4\times$ \\
\cite{demmler_aby_2015,jarvinen_pilot_2019} & $\text{2PC}_\text{OT}$ & \(1.72 \times 10^{4}\) & $73\times$ & \(6.71 \times 10^{3}\) & $28\times$ \\
\cite{mishra_delphi_2020, rathee_cryptflow2_2020, juvekar_gazelle_2018, chen_sanns_2020} & $\text{2PC}_\text{HE}$ & \(2.61 \times 10^{3}\) & $11\times$ & \(3.69 \times 10^{2}\) & $5\times$ \\
\cite{cong_revisiting_2023} & $\text{2PC}_\text{HE}$ & \(1.05 \times 10^{6}\) & $4486\times$ & - & - \\
\cite{emp-toolkit} & $\text{2PC}_\text{GC}$ & \(4.96 \times 10^{3}\) & $21\times$ & \(2.12 \times 10^{3}\) & $9\times$ \\
Crypto-$L^2$ & $\text{2PC}_\text{OT}$ & \(2.34 \times 10^{2}\) & - & \(7.40 \times 10^{1}\) & - \\
\midrule
\multicolumn{6}{l}{\textbf{Secure Clustering (LSun Dataset)}} \\
\cite{mohassel_practical_2019} & $\text{2PC}_\text{OT}$ & \(5.88 \times 10^{2}\) & $147\times$ & \(1.00 \times 10^{0}\) & $1\times$ \\
Crypto-$L^2$ & $\text{2PC}_\text{OT}$ & \(4.00 \times 10^{0}\) & - & \(1.00 \times 10^{0}\) & - \\
\midrule
\multicolumn{6}{l}{\textbf{Secure LBS Indoor (Single-Building Dataset)}} \\
\cite{jarvinen_pilot_2019} & $\text{2PC}_\text{OT}$ & \(2.90 \times 10^{4}\) & $71\times$ & \(1.25 \times 10^{4}\) & $181\times$ \\
\cite{jarvinen_pilot_2019} & $\text{2PC}_\text{GC}$ & \(7.40 \times 10^{4}\) & $181\times$ & \(1.17 \times 10^{4}\) & $170\times$ \\
Crypto-$L^2$ & $\text{2PC}_\text{OT}$ & \(4.10 \times 10^{2}\) & - & \(6.88 \times 10^{1}\) & - \\
\midrule
\multicolumn{6}{l}{\textbf{Secure Spatio-Crowdsourcing (Twitter Dataset)}} \\
\cite{han_location_2020} & $\text{SMPC}_\text{HE}$ & \(1.46 \times 10^{9}\) & $227316 \times \uparrow$ & \(1.33 \times 10^{3}\) & $0.5 \times \downarrow$ \\
Crypto-$L^2$ & $\text{2PC}_\text{OT}$ & \(6.00 \times 10^{3}\) & - & \(2.35 \times 10^{3}\) & - \\
\midrule
\multicolumn{6}{l}{\textbf{Secure Biometric Matching (AT\&T Dataset)}} \\
\cite{demmler_aby_2015} & $\text{2PC}_\text{OT}$ & \(1.19 \times 10^{3}\) & $74\times$ & \(4.91 \times 10^{2}\) & $86\times$ \\
\cite{demmler_aby_2015} & $\text{2PC}_\text{GC}$ & \(2.98 \times 10^{3}\) & $186\times$ & \(4.62 \times 10^{2}\) & $80\times$ \\
Crypto-$L^2$ & $\text{2PC}_\text{OT}$ & \(1.60 \times 10^{1}\) & - & \(5.73 \times 10^{0}\) & - \\
\midrule
\multicolumn{6}{l}{\textbf{Secure $k$-NN (SIFT Dataset, $10^6 \times 128$)}} \\
\cite{chen_sanns_2020} & $\text{2PC}_\text{HE}$ & \(5.44 \times 10^{6}\) & $13.74\times$ & \(7.54 \times 10^{5}\) & $4.96\times$ \\
\cite{cong_revisiting_2023} & $\text{2PC}_\text{HE}$ & \(2.13 \times 10^{9}\) & $5379\times$ & - & - \\
Crypto-$L^2$ & $\text{2PC}_\text{OT}$ & \(3.96 \times 10^{5}\) & - & \(1.52 \times 10^{5}\) & - \\
\midrule
\multicolumn{6}{l}{\textbf{Secure $k$-NN (Deep1B Dataset, $10^9 \times 96$)}} \\
\cite{chen_sanns_2020} & $\text{2PC}_\text{HE}$ & \(2.79 \times 10^{9}\) & $13.81\times$ & \(3.86 \times 10^{8}\) & $4.97\times$ \\
\cite{cong_revisiting_2023} & $\text{2PC}_\text{HE}$ & \(1.09 \times 10^{12}\) & $53960\times$ & - & - \\
Crypto-$L^2$ & $\text{2PC}_\text{OT}$ & \(2.02 \times 10^{8}\) & - & \(7.76 \times 10^{7}\) & - \\
\midrule
\multicolumn{6}{l}{\textbf{Secure $k$-NN (Amazon Dataset, $2^{20} \times 96$)}} \\
\cite{chen_sanns_2020} & $\text{2PC}_\text{HE}$ & \(2.74 \times 10^{6}\) & $15.22\times$ & \(3.77 \times 10^{5}\) & $4.97\times$ \\
\cite{cong_revisiting_2023} & $\text{2PC}_\text{HE}$ & \(1.06 \times 10^{9}\) & $5889\times$ & - & - \\
Crypto-$L^2$ & $\text{2PC}_\text{OT}$ & \(1.80 \times 10^{5}\) & - & \(7.58 \times 10^{4}\) & - \\

\bottomrule
\end{NiceTabular}
\label{tab:merged_l2_all_final}
\end{table}

\noindent\textbf{Secure ML Tasks.}
Many ML tasks rely on the $L^2$-norm. Mohassel et al.~\cite{mohassel_practical_2019} use it in their main clustering algorithm, and Crypto-$L^2$ achieves $271\times$ faster runtime and $4\times$ less communication under the same settings. We also compare against SAANS~\cite{chen_sanns_2020} and Cong et al.~\cite{cong_revisiting_2023} for secure $k$-NN. SAANS adopts the same homomorphic multiplication scheme as~\cite{juvekar_gazelle_2018, mishra_delphi_2020, rathee_cryptflow2_2020}; re-implementing it, we observe $11\times$ faster runtime and $5\times$ lower communication. Against Cong et al., Crypto-$L^2$ is $4486\times$ faster for vectors of dimension $2^{16}$. We further evaluate Crypto-$L^2$ on real datasets. For clustering, we compare with Mohassel et al.~\cite{mohassel_practical_2019} using the Lsun dataset~\cite{ultsch_fundamental_2020}. For $k$-NN, we compare with SAANS and Cong et al. on SIFT~\cite{lowe_object_1999}, Deep1B~\cite{yandex_efficient_2016}, and Amazon~\cite{mcauley_image-based_2015} datasets.

\vspace{0.05in}

\noindent\textbf{Secure LBS.}
We benchmark Crypto-$L^2$ against PILOT~\cite{jarvinen_pilot_2019} and Han et al.~\cite{han_location_2020}, both of which use the Euclidean distance. For PILOT, we re-implement their OT- and GC-based protocols using ABY~\cite{demmler_aby_2015}. Crypto-$L^2$ improves over PILOT by $73\times$ and $28\times$ (runtime and communication) on the OT-based version, and by $170\times$ and $26\times$ on the GC-based one. For Han et al., based on their reported complexity, we find that while their communication is lower ($0.35\times$ ours), their runtime is nearly $10^5\times$ higher. We also evaluate Crypto-$L^2$ for indoor localization using the single-building dataset~\cite{montoliu_indoorloc_2017,yang_death_2018,jarvinen_pilot_2019} and for spatial crowdsourcing with the Twitter dataset~\cite{yuan_priradar_2020}.

\vspace{0.05in}

\noindent\textbf{Secure Biometric Matching.}
We benchmark Crypto-$L^2$ against the ABY framework~\cite{demmler_aby_2015}, which also uses the Euclidean distance. Our results match the improvements seen over PILOT. We omit GSHADE~\cite{bringer_gshade_2014} since it is less efficient than ABY. We also evaluate on the AT\&T dataset~\cite{samaria_parameterisation_1994}, which contains 400 face images of size $92 \times 112$. Following~\cite{bringer_gshade_2014, sadeghi_ecient_2009}, we project the data onto 12 Eigenfaces, yielding a $400 \times 12$ matrix for evaluation.

\vspace{0.05in}

\noindent\textbf{Other Benchmarks.}
Using the EMP-Toolkit~\cite{emp-toolkit}, we implement a GC-based $L^2$-norm protocol. Crypto-$L^2$ outperforms it by $21\times$ in runtime and $9\times$ in communication.

\subsection{Evaluations for Crypto-$L^{\infty}$}
Next, we benchmark the communication cost and runtime of Crypto-$L^{\infty}$ against SOTA works (Table~\ref{tab:linfty_merged_all}).

\begin{table}[!ht]
\setlength{\tabcolsep}{2pt}
\small
\centering
\caption{Comparison of Crypto-$L^\infty$ and prior works across various secure computation tasks.}
\begin{NiceTabular}{@{}c c c c c c@{}}
\toprule
Systems & Protocol & Runtime (ms) & Speedup & Comm. (MB) & Improv. \\
\midrule\multicolumn{6}{l}{\textbf{Secure $L^{\infty}$ Norm (Synthetic $2^{16} \times 32$ Dataset)}} \\
\cite{demmler_aby_2015} & $\text{2PC}_\text{GC}$ & \(6.35 \times 10^4\) & $92\times$ & \(6.95 \times 10^2\) & $11\times$ \\
\cite{han_location_2020} & $\text{SMPC}_\text{HE}$ & \(6.60 \times 10^7\) & $10^5\times$ & \(6.40 \times 10^2\) & $19\times$ \\
\cite{mohassel_practical_2019} & $\text{2PC}_\text{GC}$ & \(2.91 \times 10^4\) & $42\times$ & \(1.32 \times 10^3\) & $21\times$ \\
\cite{demmler_aby_2015} & $\text{2PC}_\text{GMW}$ & \(1.01 \times 10^5\) & $146\times$ & \(9.87 \times 10^2\) & $16\times$ \\
\cite{mishra_delphi_2020, mohassel_aby3_2018} & $\text{2PC}_\text{GC}$ & \(9.81 \times 10^3\) & $14\times$ & \(6.69 \times 10^3\) & $106\times$ \\
\cite{huang_cheetah_2022, rathee_cryptflow2_2020} & $\text{2PC}_\text{OT}$ & \(4.62 \times 10^5\) & $671\times$ & \(3.13 \times 10^3\) & $49\times$ \\
Crypto-$L^{\infty}$ & $\text{2PC}_\text{OT}$ & \(6.88 \times 10^2\) & - & \(6.32 \times 10^1\) & - \\
\midrule
\multicolumn{6}{l}{\textbf{Secure Clustering (LSun Dataset)}} \\
\cite{mohassel_practical_2019} & $\text{2PC}_\text{GC}$ & \(3.13 \times 10^2\) & $7\times$ & \(8.00 \times 10^0\) & $8\times$ \\
Crypto-$L^{\infty}$ & $\text{2PC}_\text{OT}$ & \(4.20 \times 10^1\) & - & \(1.00 \times 10^0\) & - \\
\midrule
\multicolumn{6}{l}{\textbf{Secure Spatio-Crowdsourcing (Twitter Dataset)}} \\
\cite{han_location_2020} & $\text{SMPC}_\text{HE}$ & \(6.98 \times 10^7\) & $6348\times$ & \(4.28 \times 10^4\) & $40\times$ \\
Crypto-$L^{\infty}$ & $\text{2PC}_\text{OT}$ & \(1.08 \times 10^4\) & - & \(1.06 \times 10^3\) & - \\
\bottomrule
\end{NiceTabular}
\label{tab:linfty_merged_all}
\end{table}

\vspace{0.05in}

\noindent\textbf{Secure Clustering.}
We benchmark Crypto-$L^{\infty}$ against the $L^{\infty}$-based clustering protocol of Mohassel et al.~\cite{mohassel_practical_2019}, achieving over $42\times$ and $21\times$ reductions in runtime and communication, respectively. We further compare both systems on the Lsun dataset~\cite{ultsch_fundamental_2020}.

\vspace{0.05in}

\noindent\textbf{Secure LBS.}
We evaluate Crypto-$L^{\infty}$ against Han et al.’s crowdsourcing protocol~\cite{han_location_2020}, which uses the $L^{\infty}$-norm for secure LBS. Crypto-$L^{\infty}$ achieves up to $10^5\times$ faster execution and $10\times$ lower communication. We also simulate their spatio-crowdsourcing setting using the Twitter dataset~\cite{yuan_priradar_2020}.

\vspace{0.05in}

\noindent\textbf{Other Benchmarks.}
We compare Crypto-$L^\infty$ with SOTA inference systems including Cheetah~\cite{huang_cheetah_2022} and CrypTFlow2~\cite{rathee_cryptflow2_2020}, which share a maxpooling subroutine based on the max protocol—also central to $L^\infty$. Crypto-$L^\infty$ achieves up to $670\times$ faster runtime and $49\times$ lower communication. We also re-implement Crypto-$L^\infty$ using GC backends from DELPHI~\cite{mishra_delphi_2020} and ABY3~\cite{mohassel_aby3_2018}, resulting in $14\times$ and $106\times$ improvements, respectively. Finally, we benchmark against the ABY framework~\cite{demmler_aby_2015}, re-implementing the $L^\infty$-norm protocol using both GMW- and GC-based modes. Crypto-$L^\infty$ outperforms ABY by at least $92\times$ in runtime and $10\times$ in communication.

\section{Secure AdderNet vs. Secure CNN} \label{eval:addernet}
In Table~\ref{tab:addernet}, we report baseline plaintext accuracy for ResNet32 and MiniONN using either convolution or adder layers. As shown, convolution-based models require significantly more multiplications than adder-based ones, while achieving comparable accuracy. The multiplication counts exclude constant multiplications, which are locally computable and irrelevant in secure settings.

\begin{table}[!h]
\centering
\setlength{\tabcolsep}{3pt}
\caption{Accuracy of MiniONN and ResNet32, implemented with and without adder layers (CIFAR-10).}
\small
\begin{NiceTabular}{@{}cccc@{}}
\toprule
\multicolumn{1}{c}{Method} & \multicolumn{1}{c}{Model} & \multicolumn{1}{c}{Accuracy} & \multicolumn{1}{c}{No. of Multiplications} \\ \midrule
CNN & \multirow{2}{*}{MiniONN} & 0.9041 & 61,080,576 \\ 
AdderNet &  & 0.8981 & 174,080 \\ \midrule 
CNN & \multirow{2}{*}{Resnet32} & 0.9111 & 239,996,992 \\
AdderNet &  & 0.8756 & 1,445,952 \\ \midrule
\end{NiceTabular}
\label{tab:addernet}
\end{table}

Although AdderNet is designed for efficiency, no stable GPU-optimized implementation currently exists, making direct runtime comparisons with highly optimized convolutions infeasible. Nonetheless, the results demonstrate that AdderNet can serve as a practical substitute for CNNs without significant accuracy loss.

\begin{table}[!h]
\centering
\setlength{\tabcolsep}{2pt}
\caption{Secure AdderNet (Crypto-$L^1$) vs. CrypTFlow2 \cite{rathee_cryptflow2_2020} and Cheetah \cite{huang_cheetah_2022}, for an input length of $32$ bits over $2^8$, $2^{14}$, and $2^{16}$ operations.}
\small
\begin{NiceTabular}{@{}ccccc@{}}
\toprule
Dimension & Operation & Protocol & \multicolumn{1}{c}{Runtime (ms)} & \multicolumn{1}{c}{Comm. (MB)} \\ \midrule
\multirow{2}{*}{$2^{8}$} & Conv & \cite{rathee_cryptflow2_2020,huang_cheetah_2022} & $2$ & $0.29$ \\
 & Adder & Crypto-$L^1$ & $5$ & $0.13$ \\ \midrule
\multirow{2}{*}{$2^{14}$} & Conv & \cite{rathee_cryptflow2_2020,huang_cheetah_2022} & $59$ & $18.5$ \\
 & Adder & Crypto-$L^1$ & $83$ & $8.37$ \\ \midrule
\multirow{2}{*}{$2^{16}$} & Conv & \cite{rathee_cryptflow2_2020,huang_cheetah_2022} & $234$ & $74$ \\
& Adder & Crypto-$L^1$ & $324$ & $32.47$ \\\bottomrule
\end{NiceTabular}
\label{tab:adder_iknp}
\end{table}

In Table~\ref{tab:adder_iknp} (Appendix~\ref{tab:adder_iknp}), we present a comparison between our secure Adder operation and the OT-based secure Conv operation from \cite{rathee_cryptflow2_2020, huang_cheetah_2022}\footnote{The results for SIRNN~\cite{rathee_sirnn_2021} are consistent, as they all utilize the same SCI library for their IKNP-style OT-based convolution operations.}. These operations are evaluated on $n$-dimensional vectors $x$ and $y$ of 32-bit integers, where $Adder(x,y) = -|x-y|$ and $Conv(x,y) = x \cdot y$. We assess both the runtime, measured in milliseconds, and the communication overhead, in megabytes, across three different vector dimensions using syntheti data.

\begin{table}[!h]
\centering
\setlength{\tabcolsep}{2pt}
\caption{Comparison of our Secure Adder with OT-based Secure Conv from \cite{rathee_cryptflow2_2020,huang_cheetah_2022} on CIFAR-10 (ResNet-32).}
\small
\begin{NiceTabular}{@{}cccccc@{}}
\toprule
\multirow{1}{*}{Input} & \multirow{1}{*}{Kernel} & \multirow{1}{*}{(Stride,} & \multirow{2}{*}{Operation} & \multicolumn{1}{c}{Runtime} & \multicolumn{1}{c}{Comm.} \\
$h_{in}\!\times\!w_{in}\!\times\!c_{in}$ & $d\!\times\!d\!\times\!c_{out}$ & Padding) &  & (s) & (MB) \\ \midrule
\multirow{2}{*}{$32\!\times\!32\!\times\!3$} & \multirow{2}{*}{$3\!\times\!3\!\times\!16$} & \multirow{2}{*}{(1, 1)} & Conv & $1.5$ & $500$ \\
& & & Adder & $2.3$ & $224$ \\ \midrule
\multirow{2}{*}{$32\!\times\!32\!\times\!16$} & \multirow{2}{*}{$1\!\times\!1\!\times\!32$} & \multirow{2}{*}{(2, 0)} & Conv & $0.5$ & $148$ \\
& & & Adder & $0.6$ & $60$ \\ \midrule
\multirow{2}{*}{$16\!\times\!16\!\times\!32$} & \multirow{2}{*}{$3\!\times\!3\!\times\!64$} & \multirow{2}{*}{(2, 1)} & Conv & $4$ & $1332$ \\
& & & Adder & $6$ & $597$ \\ \midrule
\multirow{2}{*}{$8\!\times\!8\!\times\!64$} & \multirow{2}{*}{$3\!\times\!3\!\times\!64$} & \multirow{2}{*}{(1, 1)} & Conv & $7.2$ & $2664$ \\
& & & Adder& $12$ & $1000$ \\ \bottomrule
\end{NiceTabular}
\label{tab:adder_vs_vonc_cifar10}
\end{table}

We extend our comparison to real-world data by re-evaluating both operations on the CIFAR-10 dataset using the ResNet-32 architecture (Table~\ref{tab:adder_vs_vonc_cifar10}). In this setting, the Adder operation achieves up to $3\times$ lower communication than Conv, albeit with slightly slower runtime—highlighting a trade-off between efficiency and speed.

\begin{figure}[!h]
\centering
\subfigure[Communication \label{fig:adder_vs_conv_comm}]{
\includegraphics[width=0.4625\columnwidth]{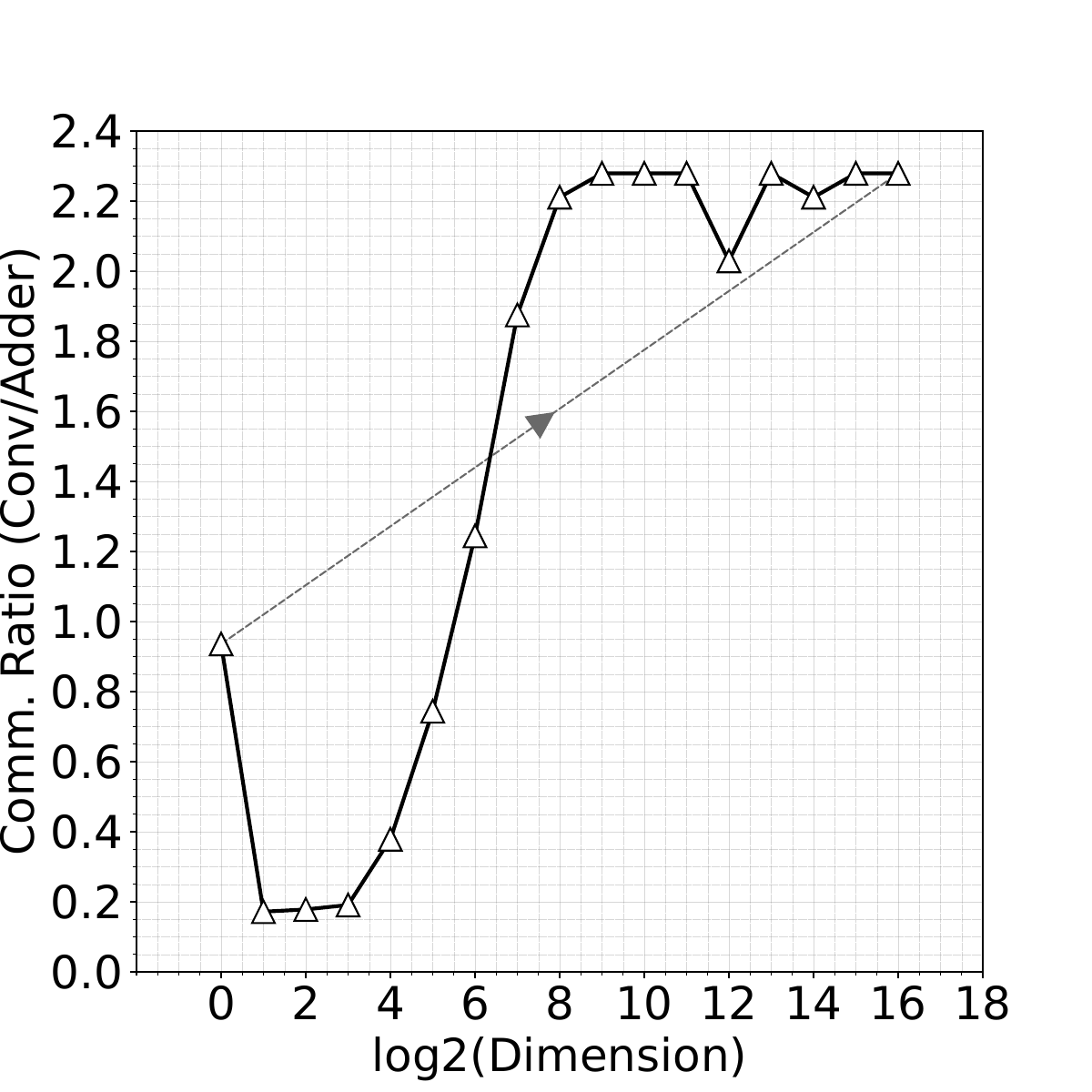}}
\subfigure[Runtime \label{fig:adder_vs_conv_runtime}]{
\includegraphics[width=0.4625\columnwidth]{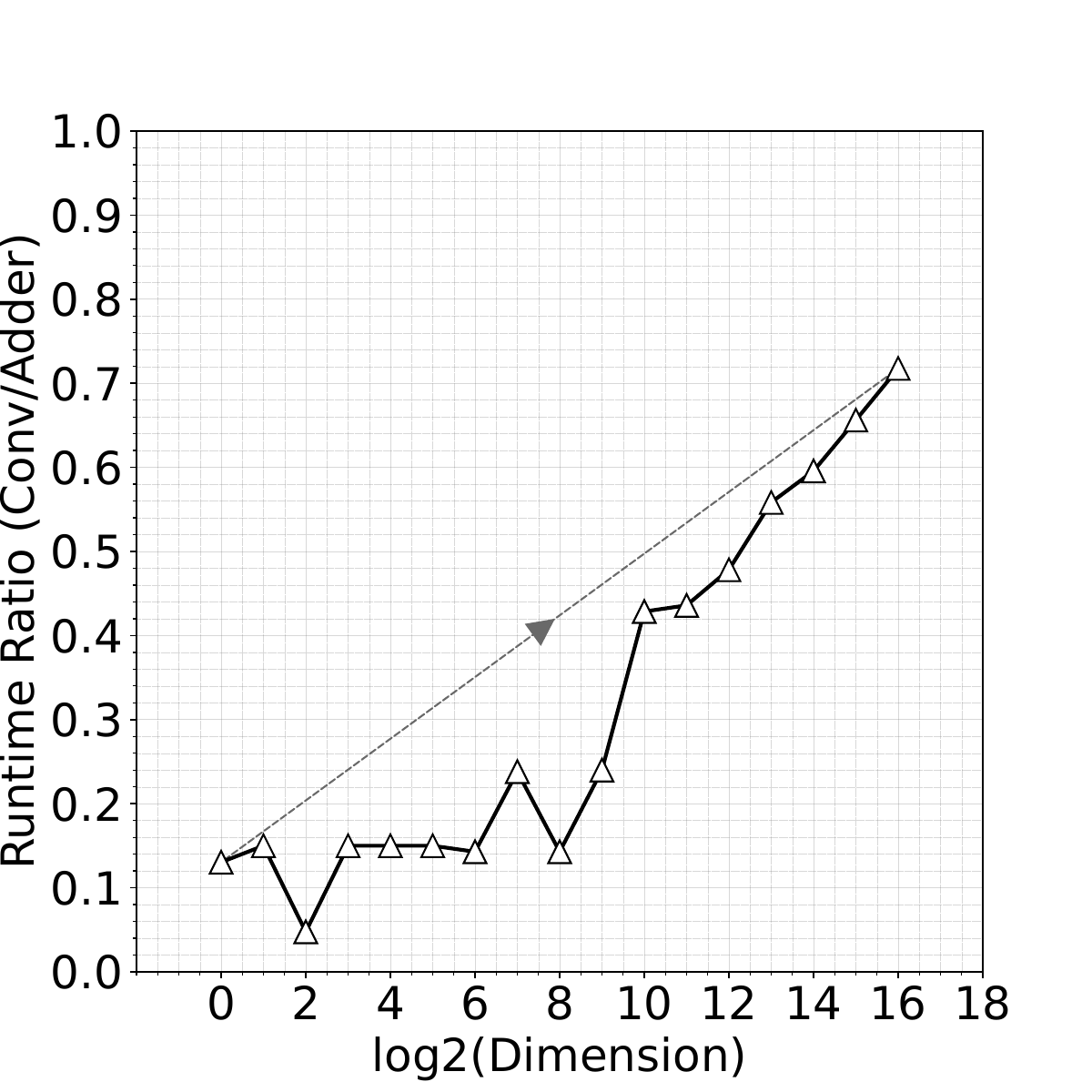}}
\caption{Adder vs. Conv operations across various input dimensions.}\vspace{-0.1in}
\label{fig:adder_vs_conv}
\end{figure}

To further analyze this trade-off, we evaluate secure Adder and Conv operations across dimensions from $1$ to $2^{16}$, averaging runtime and communication ratios over 10 runs per dimension. Results are reported in Figure~\ref{fig:adder_vs_conv}. As shown in Figure~\ref{fig:adder_vs_conv_comm}, Adder reduces communication significantly for dimensions above $2^6$, with a peak improvement of $\sim 2.4\times$, though it is marginally worse in lower dimensions. Figure~\ref{fig:adder_vs_conv_runtime} shows that Conv is slightly faster overall, but the gap narrows with increasing dimension, reaching near parity beyond $2^{14}$.

\section{Related Work} \label{sec:related}
A series of works on secure two-party clustering~\cite{jaschke_unsupervised_2019, bunn_secure_2007, jagannathan_privacy-preserving_2005, mohassel_practical_2019} have proposed protocols for securely computing distances between data points and cluster centroids. Most notably, \cite{bunn_secure_2007, jagannathan_privacy-preserving_2005, mohassel_practical_2019} use the Euclidean distance as their primary metric. In contrast, Jäschke et al.~\cite{jaschke_unsupervised_2019} employ the Manhattan distance to avoid the cost of multiplications and the increased bit length introduced by Euclidean computations. Earlier works~\cite{bunn_secure_2007, jagannathan_privacy-preserving_2005} rely on semantically secure homomorphic encryption to compute multiplications securely. Mohassel et al.~\cite{mohassel_practical_2019} instead adopt OT-based multiplication triples~\cite{demmler_aby_2015}, enhanced with $\binom{N}{1}\text{-OT}_{\ell}$ from~\cite{dessouky_pushing_2018}, and further reduce overhead via a sequential amortized multiplication protocol that amortizes one OT across $T$ multiplications (where $T$ is the number of clusters). They also benchmarked clustering with $L^1$ and $L^\infty$ norms and concluded that $L^2$-norm outperforms both in efficiency.

Similarly, secure biometric matching protocols~\cite{demmler_aby_2015, bringer_gshade_2014, hutchison_automatic_2014, bringer_privacy-preserving_2013, blanton_secure_2010} use the squared Euclidean distance to compare biosamples and identify the closest match. GSHADE~\cite{bringer_gshade_2014} used OT extensions for secure biometric identification. The ABY framework~\cite{demmler_aby_2015} improved upon this with a mixed-protocol design combining Arithmetic, Boolean, and Yao sharings.

Privacy-preserving distance computation is also central to location-based services (LBSs)~\cite{han_location_2020, huang_secure_2016}. Han et al.~\cite{han_location_2020} propose secure distance protocols for spatial crowdsourcing (SC), where a server coordinates assignments without learning sensitive location data, and a trusted party handles key distribution. A core building block in LBS is the $k$-nearest neighbors ($k$-NN) algorithm, typically based on secure Euclidean distance~\cite{chen_sanns_2020, cong_revisiting_2023, zuber_efficient_2021}. In contrast, Shaul et al.~\cite{shaul_secure_2020} propose a $k$-ish NN variant using the $L^1$-norm. Xie et al.~\cite{XieHW19,XieHW20} propose cryptographic protocol to approximate the absoluate difference of power loads using the $L^1$-norm. Chen et al.~\cite{chen_sanns_2020}, along with~\cite{cong_revisiting_2023, zuber_efficient_2021}, employ leveled and fully homomorphic encryption~\cite{fan_somewhat_2012, brakerski_leveled_2012, brakerski_fully_2012} to compute squared Euclidean distances in $k$-NN applications.

\section{Conclusion}

In this paper, we present \sys{}, a versatile framework for secure $L^p$-norm computation in 2PC settings, applicable to various use cases. Our extensive evaluations demonstrate that \sys{} consistently outperforms previous works on $L^p$-norms, particularly \cite{mohassel_practical_2019}, across diverse real-world scenarios and datasets.
Additionally, by incorporating the $L^1$-norm into machine learning inference, we develop $L^p$-norm-based CNN models (e.g., AdderNet) that achieve improved communication efficiency compared to SOTA systems \cite{rathee_cryptflow2_2020, huang_cheetah_2022}, while maintaining similar runtime and accuracy.

\section*{Acknowledgement}
This work is partially supported by the National Science Foundation (NSF) under Grants No. CNS-2302689, CNS-2308730, CNS-2319277, CNS-2432533, CNS-2034615, CMMI-2326340, CMMI-2326341, as well as by a Cisco Research Award and a Synchrony Fellowship.

\appendix
\subsection{Secure 2PC DNN Inference} \label{appendix:cnn}
A DNN takes as input a tensor $X$ and processes it through a sequence of linear and non-linear layers to classify it into one of the potential classes. For simplicity, we can conceptualize a DNN as a sequence of functionalities $f_i$, where each $f_i$ represents the function evaluated in the $i$-th layer. Accordingly, we can express the output $z$ of each layer as $z = f_i(f_{i-1}(\ldots(f_1(X_1, W_1), \ldots), W_{i-1}), W_i)$, where $X_i$ and $W_i$ represent the input tensor and the weights of the $i$-th layer, respectively. In convolutional neural networks (CNNs), which are a subtype of DNNs, some of these functionalities, $f_i$, are performed through convolution operations. Convolutions measure the similarity between the input tensors and the weights (or filters $F_i$) by computing the element-wise product of them as the filter, sized $d \times d$, slides across the input. The results of these products are then summed up at each position of the slide. In secure 2PC inference tasks \cite{DBLP:conf/codaspy/LiuAWLBZH25,DBLP:conf/arxiv/LeeABLH25,mishra_delphi_2020,rathee_cryptflow2_2020}, two parties, $P_0$ and $P_1$ hold $X$ and $W$, respectively. They securely compute the inference result, while ensuring that $P_0$ does not learn anything about $P_1$'s proprietary weights $W$, and $P_1$ does not learn anything about $P_0$'s input $X$.

\subsection{Building Blocks: Extended}
\subsubsection{Correctness for $\mathcal{F}_{\text{ABS\_MUX}}^{\ell}$ (Alg.~\ref{alg:abs_mux})} \label{sec:correct_abs_mux} In order to prove the correctness of our $\mathcal{F}_{\text{ABS\_MUX}}^{\ell}$ functionality, we need to first transform the XOR operation, originally applied to boolean shares $\bshare{s}_b$, into its equivalent operation on their corresponding arithmetic values $s_b$. This is achieved through the use of a generalized XOR operation that can be defined in any field \cite{rotaru_marbled_2019}. Specifically, we have the following:
\begin{equation} \label{eq:arth_xor}
	\bshare{s}_0 \oplus \bshare{s}_1 = s_0 + s_1 - 2 \cdot s_0 \cdot s_1,
\end{equation}
where $s_b$ denotes the $\ell$-bit arithmetic equivalent of $\bshare{s}_b$. It is straightforward for a party $P_b$ to convert from $\bshare{s}_b$ to $s_b$ without incurring extra cost. They can simply set the least significant bit of $s_b$ to $\bshare{s}_b$, and all other bits to $0$s. Due to this trivial conversion, we leave $\bshare{s}_b$ as is in Alg.~\ref{alg:abs_mux}.

As noted in Section~\ref{sec:multiplexer_abs}, the $\mathcal{F}_{\text{ABS\_MUX}}^{\ell}$ functionality, given an $\ell$-bit integer $x$ and selection bit $s$, computes $-x$ if $s=1$ and $x$ otherwise. Algebraically, this is equivalent to computing:
{
\small
\begin{align*}
	 & \phantom{{}={}} (1 - 2 \cdot s) \cdot x                                                                                                   \\
	 & = (1 - 2 \cdot (\bshare{s}_0 \oplus \bshare{s}_1)) \cdot (\ashare{x}{L}_0 + \ashare{x}{L}_1)                                              \\
	 & = (1 - 2 \cdot (s_0 + s_1 - 2 \cdot s_0 \cdot s_1)) \cdot (\ashare{x}{L}_0 + \ashare{x}{L}_1)                                             \\
	 & = \ashare{x}{L}_0 - 2 \cdot (s_0 \cdot \ashare{x}{L}_0 + s_1 \cdot (\ashare{x}{L}_0 - 2 \cdot s_0 \cdot \ashare{x}{L}_0))                 \\
	 & \phantom{{}={}} + \ashare{x}{L}_1 - 2 \cdot (s_1 \cdot \ashare{x}{L}_1 + s_0 \cdot (\ashare{x}{L}_1 - 2 \cdot s_1 \cdot \ashare{x}{L}_1)) \\
	 & = \ashare{x}{L}_0 - 2 \cdot (s_0 \cdot \ashare{x}{L}_0 + s_0 \cdot (\ashare{x}{L}_1 - 2 \cdot s_1 \cdot \ashare{x}{L}_1))                 \\
	 & \phantom{{}={}} + \ashare{x}{L}_1 - 2 \cdot (s_1 \cdot \ashare{x}{L}_1 + s_1 \cdot (\ashare{x}{L}_0 - 2 \cdot s_0 \cdot \ashare{x}{L}_0)) \\
	 & = w_0 - 2 \cdot y_0 + w_1 - 2 \cdot y_1                                                                                                   \\
	 & = \ashare{z}{L}_0 + \ashare{z}{L}_1 \qed
\end{align*}
}

\normalsize

\subsubsection{Base Multiplexer} \label{appendix:base_mux}
In this work, we define a ``base multiplexer'' as a secure functionality $\func{Base\_MUX}{\ell}$, which takes shares of an $\ell$-bit integer $x$ and a selector $s$, and returns shares of $x$ if $s=1$ and $0$ otherwise. We design $\func{Base\_MUX}{\ell}$ (see Alg.~\ref{alg:base_mux}) to realize this functionality, and employ an approach similar to \cite{rathee_sirnn_2021}, which improves over the multiplexer used in \cite{rathee_cryptflow2_2020} by leveraging $\binom{2}{1}\text{-COT}_\ell$ instead of $\binom{2}{1}\text{-OT}_\ell$, and reduces communication costs.

\begin{algorithm}
	\small
	\caption{Base Multiplexer, $\mathcal{F}^{\ell}_{\text{Base\_MUX}}$}\label{alg:base_mux}
	\begin{algorithmic}[1]
		\Input{For $b \in \{0,1\}$, $P_b$ holds $\ashare{x}{L}_b$ and $\bshare{s}_b$.}
		\Output{For $b \in \{0, 1\}$, $P_b$ learns $\ashare{z}{L}_b$ s.t. $z = x$ if $s = 1$, else \hspace*{-0.8mm} $z = 0$.}
		\State \multiline{%
			For $b \in \{0, 1\}$, $P_b$ chooses $r_b \overset{\$}{\leftarrow} \mathbb{Z}_{L}$ and computes $w_b = \ashare{x}{L}_b \cdot \bshare{s}_b - r_b$.
		}
		\State \multiline{%
			$P_0$ and $P_1$ invoke an instance of ${2 \choose 1}\text{-COT}_\ell$ where $P_0$ is the sender with correlation function $\rho(r_0) = r_0 + \ashare{x}{L}_0 - 2 \cdot \bshare{s}_0 \cdot \ashare{x}{L}_0$ and $P_1$ is the receiver with input $\bshare{s}_1$. Party $P_1$ learns $y_1 = r_0 + \bshare{s}_1(\ashare{x}{L}_0 - 2 \cdot \bshare{s}_0 \cdot \ashare{x}{L}_0)$.}
		\State \multiline{%
			$P_0$ and $P_1$ invoke an instance of ${2 \choose 1}\text{-COT}_\ell$ where $P_1$ is the sender with correlation function $\rho(r_1) = r_1 + \ashare{x}{L}_1 - 2 \cdot \bshare{s}_1 \cdot \ashare{x}{L}_1$ and $P_0$ is the receiver with input $\bshare{s}_0$. Party $P_0$ learns $y_0 = r_1 + \bshare{s}_0(\ashare{x}{L}_1 - 2 \cdot \bshare{s}_1 \cdot \ashare{x}{L}_1)$.}
		\State For $b \in \{0, 1\}$, $P_b$ outputs $\ashare{z}{L}_b =  w_b + y_b$.
	\end{algorithmic}
\end{algorithm}

\noindent \textit{Security Analysis and Communication Complexity.} Our $\mathcal{F}_{\text{Base\_MUX}}^{\ell}$ functionality requires two calls to $\binom{2}{1}\text{-COT}_{\ell}$, which is equal to $2(\lambda+\ell)$, whereas in \cite{rathee_cryptflow2_2020} they use two calls to $\binom{2}{1}\text{-OT}_{\ell}$, which has a cost of $2(\lambda+2\ell)$. Security follows trivially in $\binom{2}{1}\text{-COT}_{\ell}$ hybrid.



\normalsize

\subsubsection{Generalized Multiplexer} \label{appendix:gen_mux}
In this section, we design a protocol for a multiplexer that selects between arbitrary inputs ($a$ and $a'$). Similar to the ABS multiplexer (Section~\ref{sec:secure_absolute}), we want to compute $s \cdot a + (1-s) \cdot a' = a' + s \cdot (a-a')$. Note that in Alg.~\ref{alg:abs_mux}, we had that $a = -x$ and $a' = x$; thus, this expression simplified to $(1 - 2 \cdot s) \cdot x$. In Alg.~\ref{alg:gen_mux}, we show how we can construct this functionality $\mathcal{F}^{\ell}_{\text{MUX}}$ based on the base multiplexer without incurring additional costs.


\noindent \textit{Security Analysis and Communication Complexity.} In Alg.~\ref{alg:gen_mux}, we only communicate in the protocol for $\func{\text{Base\_MUX}}{\ell}$, which costs $2(\lambda+\ell)$. Security follows from the security of $\func{\text{Base\_MUX}}{\ell}$.

\begin{algorithm}
	\small
	\caption{Generalized Multiplexer, $\mathcal{F}^{\ell}_{\text{MUX}}$}\label{alg:gen_mux}
	\begin{algorithmic}[1]
		\Input{For $b \in \{0,1\}$, $P_b$ holds $\ashare{a}{L}_b$, $\ashare{a'}{L}_b$, and $\bshare{s}_b$.}
		\Output{For $b \in \{0, 1\}$, $P_b$ learns $\ashare{z}{L}_b$ s.t. $z = a$ if $s = 1$, else \hspace*{-0.8mm} $z = a'$.}
		\State \multiline{%
			For $b \in \{0,1\}$, $P_b$ computes $\ashare{d}{L}_b = \ashare{a}{L}_b - \ashare{a'}{L}_b$.}
		\State \multiline{%
			$P_0$ and $P_1$ invoke an instance of $\func{Base\_MUX}{\ell}$ with inputs $(\ashare{d}{L}_0,\bshare{s}_0)$ and ($\ashare{d}{L}_1,\bshare{s}_1)$, respectively. For $b \in \{0,1\}$, $P_b$ learns $\ashare{w}{L}_b$ where $w = s \cdot d$.}
		\State \multiline{%
			For $b \in \{0,1\}$, $P_b$ outputs $\ashare{a'}{L}_b + \ashare{w}{L}_b$.}
	\end{algorithmic}
\end{algorithm}

\bibliographystyle{IEEEtran}
\bibliography{ref}

\end{document}